\pdfoutput=1

\documentclass[a4paper,12pt]{article}
\usepackage[utf8x]{inputenc}
\usepackage{amssymb,amsmath,hyperref,graphicx}
\usepackage[a4paper, margin=1.5cm]{geometry}
\usepackage{tikz}

\newtheorem{rema}{Remark}[section]

\newcommand{\bc}{\begin{center}}
\newcommand{\ec}{\end{center}}
\def\ba#1{\begin{array}{#1}\displaystyle}
\newcommand{\ea}{\end{array}}

\newcommand{\beq}{\begin{equation}}
\newcommand{\eeq}{\end{equation}}
\newcommand{\beqa}{\begin{eqnarray}}
\newcommand{\eeqa}{\end{eqnarray}}
\newcommand{\no}{\nonumber}
\newcommand{\n}{\nonumber\\}
\newcommand{\bi}{\begin{itemize}}
\newcommand{\ei}{\end{itemize}}

\def\lt#1{\left#1}
\def\rt#1{\right#1}
\def\t#1{\tilde{#1}}

\def\frc#1#2{\frac{#1}{#2}}

\newcommand{\p}{\partial}

\newcommand{\Pexp}{{\cal P}\exp}

\newcommand{\bra}{\langle}
\newcommand{\ket}{\rangle}
\newcommand{\Z}{{\mathbb{Z}}}

\newcommand{\R}{{\mathbb{R}}}

\newcommand{\Or}{{\cal O}}

\newcommand{\Tr}{{\rm Tr}}

\newcommand{\ri}{{\rm i}}

\usetikzlibrary{positioning}
\newcommand{\insertion}[5]{\draw[fill=gray](#1+#3,#2-#4)circle(#4);\fill[black]
(#1+#3,#2-#4-.4*#5)node[scale=#5]{$p_{#3}$};}
\newcommand{\vertzero}[5]{\draw[thin, rounded
corners=2pt](#1+#4,#2)--(#1+#4,#2+0.2*#3)--(#1+#5,#2+0.2*#3)--(#1+#5,#2);}
\newcommand{\vertnonzero}[5]{\draw[thin, rounded
corners=2pt](#1+#4,#2+0.2*#3)--(#1+#5,#2+0.2*#3)--(#1+#5,#2);\draw[thin](#1+#4,
#2)--(#1+#4,#2+0.15+0.2*#3);\draw[fill=gray](#1+#4,#2+0.2*#3) circle
(0.05);\fill[black](#1+#4+.06,#2+.1+0.2*#3) node[anchor=west,scale=.75]
{$p_{#4}+p_{#5}$};}

\usetikzlibrary{decorations.markings}

\begin{document}

\begin{center}

{\Large {\bf Energy flow and fluctuations in non-equilibrium conformal field
theory on star graphs}}

\vspace{0.8cm} 

{\large \text{Benjamin Doyon${}^{\spadesuit}$,
Marianne Hoogeveen${}^{\spadesuit}$ and Denis Bernard${}^{\clubsuit}$}}

\vspace{0.2cm}
{\small ${}^{\spadesuit}$ Department of Mathematics, King's College London,
London, United Kingdom.}\\
{\small ${}^{\clubsuit}$ Laboratoire de Physique Th\'eorique de l'ENS, CNRS $\&$
Ecole Normale Sup\'erieure de Paris, France.}
\end{center}

\vspace{0.5cm} 
We consider non-equilibrium quantum steady states in conformal field theory
(CFT) on star-graph configurations, with a particular, simple connection
condition at the vertex of the graph. These steady states occur after a large
time as a result of initially thermalizing the legs of the graph at different
temperatures, and carry energy flows. Using purely Virasoro algebraic
calculations we evaluate the exact scaled cumulant generating function for
these flows. We show that this function satisfies a generalization of the usual
non-equilibrium fluctuation relations. This extends the results by two of the
authors (J. Phys. A 45: 362001, 2012; arXiv:1302.3125) to the cases of more than
two legs. It also provides an alternative derivation centered on Virasoro
algebra operators rather than local fields, hence an alternative regularization
scheme, thus confirming the validity and universality of the scaled cumulant
generating function. Our derivation shows how the usual Virasoro algebra leads,
in large 
volumes, to 
a continuous-index Virasoro algebra for which we develop diagrammatic principles,
which may be of interest in other non-equilibrium contexts as well. Finally, our
results shed light on the Poisson process interpretation of the long-time energy
transfer in CFT.
\medskip

\hfill \today
\medskip


\section{Introduction}

There is currently great interest in the thermodynamics of quantum systems out
of equilibrium. This has to do in part with the general need to find a framework
that extends equilibrium thermodynamics to far-from-equilibrium situations
\cite{zwanzig2001nonequilibrium}, and in part with the recent experimental
advances which make it possible to prepare quantum systems in non-equilibrium
states in a controlled way, and verify fluctuation relations experimentally
\cite{utsumi2010bidirectional,nakamura2010nonequilibrium,
nakamura2011fluctuation,sanchez2012detection,saira2012test}. In the context of
non-equilibrium quantum steady states, where there is transfer of energy,
charge, particles, etc., one of the objects of interest is the scaled
cumulant generating function (the Legendre transform of the large-deviation
function), see for instance the review \cite{esposito2009nonequilibrium}. This characterizes the fluctuations of these transfers measured over a large time period,
and encodes many properties of their non-equilibrium statistics. Calculating it
exactly in model systems and 
obtaining the associated fluctuation relations are important steps in developing
the general theory of non-equilibrium steady states.

Recently, a step in this direction was achieved by calculating exactly the
scaled cumulant generating function in non-equilibrium conformal field theory
(NECFT) \cite{bernard2012energy,bernard2013non}, for energy and charge transfer. The setup consisted of two systems (acting as baths), initially thermalized at different
temperatures and chemical potentials, that were connected to form a homogeneous system,
so that a steady state is established at late times. Besides deriving the
non-equilibrium density matrix and proving an exact nontrivial formula for the
cumulant generating function in this relatively simple situation of NECFT, two
claims were made. Firstly, it was claimed (with justifications) in \cite{bernard2012energy,bernard2013non} that
this formula is {\em universal}. That is, it gives the correct scaled
cumulant generating function for any quantum critical system\footnote{With
dynamical critical exponent $z=1$.}, prepared in the way described above, at small temperatures
and chemical potentials.\footnote{There are subtleties regarding the interplay between the large-$t$ limit and the scaling limit. These are discussed in \cite{bernard2013non}.} This is a nontrivial statement, meaning that for the
first time, 
the exact long-time fluctuation statistics were obtained for both energy and
charge transfer for many families of quantum systems, even with very strong
interactions. The only parameter needed, that encodes the interaction, is the
central charge. The NECFT universal average current was recently confirmed
numerically \cite{Karrasch2012non} in a particular critical quantum model. It was also observed
in \cite{bernard2012energy} that the long-time energy transfer statistics is purely Poissonian,
leading to the claim that the natural right-moving and left-moving energy
carriers behave over a long time like ensembles of Poissonian particles.

The objective of the present paper is two-fold. Firstly, we generalize the
results of \cite{bernard2012energy} to non-equilibrium energy flows amongst any
number of baths, connected to each other at one point in a star-graph
configuration. We obtain the density matrix and derive the exact universal
scaled cumulant generating function (which we will sometimes refer to as the
full counting statistics) in NECFT under a simple connection condition at the
vertex of the graph. Our results are a simple generalization of those of
\cite{bernard2012energy}, and we find a generalization of the standard fluctuation relations to
this many-leg case. There has been a lot of activity in the study of such
``quantum graphs'' (see for instance \cite{kuchment2008} for a review, and
\cite{Nayak1999resonant,Oshikawa2006Junction,Hou2009Corner,
Agarwal2009enhancement,Safi2001partition, Beri2012Topological, Tsvelik2013Majorana, Crampe2013Quantum} for condensed matter applications).
Recently, certain quantum field theory methods and concepts have been introduced (see e.g.
\cite{fridean2005entropy,PhysRevB.74.045322,bellazzini2006quantum,
bellazzini2007bosonization,bellazzini2008quantum,ragoucy2009quantum,
Rahmani2012general}) and studies have looked at far-from-equilibrium quantities
in various types of quantum graphs
\cite{Safi2001partition,mintchev2011non,caudrelier2013quantum,
mintchev2013luttinger}. However, to our knowledge no exact cumulant generating
function has been derived yet, and most of these studies are restricted to free
propagation in the legs of the graph (e.g.~Luttinger liquids), concentrating
on the nontrivial scattering at vertices (see for instance the methods developed
in \cite{caudrelier2009direct}). By contrast our results apply to general CFT in
the legs, possibly representing strongly interacting critical systems, albeit
with simple vertex scattering.

Secondly, we develop a new method for deriving the scaled cumulant generating
function. Some of the problems in using conformal field theory (CFT) methods in
NECFT are that (1) in order to describe baths, the systems must be of infinite
length, and (2) there is (yet) no simple Euclidean-space geometry to
re-interpret out of equilibrium states and their fluctuations, so that we have a
truly real-time problem where the quantization scheme is fixed {\em a priori}.
This hampers the use of standard techniques based on the Virasoro
algebra and its representations, which usually require the choice of a
quantization scheme where space is compact. This problem was circumvented in
\cite{bernard2012energy} by concentrating on the algebra of local fields on the line. Here we
attack the problem directly, and develop a method to calculate quantum averages
of Virasoro generators in the limit where the length of the system goes to
infinity. To that end, we study a continuous-index Virasoro algebra and develop
associated 
diagrammatic principles. This algebraic method could find applications in other
contexts, for instance in the study of non-equilibrium steady states in NECFT
with nontrivial impurities (impurities are usually described in terms of the
Virasoro algebra), or in quantum quenches.

Besides obtaining the non-equilibrium density matrix and deriving the exact
scaled cumulant generating function in NECFT with the present simple
star-graph configuration, our results provide further evidence for the two
claims mentioned above. Indeed, within our new calculation method, there is a
natural UV regularization scheme, closely linked to the usual UV regularization
of quantum field theory, but different from that used in \cite{bernard2013non}. Since our
results agree with those of \cite{bernard2012energy,bernard2013non} in the two-leg case, this provides further confirmation of the
expected independence from the regularization scheme, and consequently universality (although it is not a full proof or a full analysis of the effect of irrelevant operators). Also,
our cumulant generating function has a clear Poisson-process interpretation
further confirming the interpretation of \cite{bernard2012energy}.

The paper is organized as follows. In Section \ref{sect:1}, we describe precisely
the system under consideration and obtain the non-equilibrium density matrix. In
Section \ref{sect:2}, we describe our main results and discuss their meaning. In
Section \ref{sect:3}, we develop some aspects of the continuous Virasoro algebra;
in particular, we study the diagrams used to evaluate averages. In Section
\ref{sect:4} we evaluate the full counting statistics. Finally, in Section
\ref{sect:conclu}, we provide concluding remarks.


\section{A nonequilibrium steady state on a star graph}
\label{sect:1}

Consider a number $N$ of identical, decoupled one-dimensional quantum systems,
each of length $R/2$ and at criticality, thermalized at inverse temperatures
$\beta_1,\ldots,\beta_N$. Assume that the temperatures are small as compared
with the microscopic energy scale $J$ (the typical energy of a link in a quantum
chain, for instance), $k_BT\ll J$. Then the physics of each quantum system can
be described by a conformal field theory (CFT), with some central charge $c$. In
particular, the boundaries are also conformal.
Since the systems are described by a CFT, the energy and momentum densities (and
their descendants), in the bulk, separate into left- and right-movers. We arrange the
systems radially, forming the legs (or edges) of a star graph ($N$ legs
meeting at a single vertex); then it is clearer to call the fields incoming or
outgoing, where incoming fields are the fields that move towards the vertex, and
outgoing are those that move away from the vertex. We will denote them by 
$h^{\text{in/out}}_j$, respectively, so that the energy density is $h^{\text{in}}_j
+ h^{\text{out}}_j$ and the (inward) momentum density is $h^{\text{in}}_j -
h^{\text{out}}_j$ in leg $j$. Thanks to the conformal boundary conditions, at the boundaries
of each system, we have reflective conditions for the densities
$h^{\text{in/out}}_j$ \cite{Cardy1984conformal}. For example, incoming fields become outgoing fields moving
at the same speed upon reaching the innermost boundaries of the systems, which
are located at the vertex of the star graph. 

At some time $-t_0<0$ in the far past, the independently thermalized systems are
all connected at the vertex. There are several ways of making this connection;
in this paper the connection is assumed to be made in such a way that the
incoming and the outgoing fields on one leg are connected to different legs.
More precisely, the incoming fields of the $j$-th leg, when they reach the
vertex, move into the $j+1$-th leg (mod $N$), where they become outgoing
fields. One way to think of this connection where incoming and outgoing fields
move into different legs is by considering these fields as the edge currents of
a set of $N$ very long and narrow quantum Hall slabs arranged into a star graph.

After the systems are connected, the new total system is evolved unitarily.
Then, energy will begin to flow from higher-temperatures regions to
lower-temperature regions. If we let the system evolve for a long enough time, we expect the system
to reach a steady state. More precisely, in order for the system to be in a
steady state at time $t=0$, we must take the length $R$ to infinity before we
take the limit $t_0\rightarrow\infty$. This ensures that the legs of the graph
act as reservoirs at different temperatures, and that any finite part of the
system around the vertex can be seen as an open system. Indeed, waves emanating
from the vertex will not have time to bounce back at outer boundaries, hence
will effectively be absorbed by the legs; and waves incoming to the vertex can
only come from deep inside the legs, carrying the information of the initial
thermalization. We will call this limit, $\lim_{R\gg t_0\rightarrow\infty}$, the
steady state limit, and what we mean by the system reaching a steady state is
the existence 
of the limit on averages,
\begin{equation}
\lim_{t_0\rightarrow\infty}\lim_{R\rightarrow\infty}\langle\ldots\rangle_{R,t_0}
=\langle\ldots\rangle^{\text{stat}},\label{eq:steady_state_limit}
\end{equation}
where $\bra\cdots\ket_{R,t_0}$ represents the average in the finite-$R$ star graph a length of time $t_0$ after the connection has been made. Since it is only finite parts
around the vertex that are expected to possess a steady state limit, for the
system to be said to reach a steady state we only impose this limit to exist
whenever the ellipses $\cdots$ are replaced by operators supported in finite
regions around the vertex; and in fact, we will here only look at operators in
the same Virasoro sector as that of the energy and momentum densities. We expect
there to be a steady energy current flowing, so that, in particular, the
steady-state average of the energy-current observable (the momentum density)
should be finite and nonzero. Note that any finite region around the vertex, no
matter how large, will be in a steady state in this limit. What effectively
happens in the steady state limit, is that the region in which we have a steady
energy current becomes infinitely large, and that the thermal baths are pushed
to infinity.

The distance along the any one of the $N$ legs is parametrized by $x\in[0,\frac{R}{2}]$. Before the connection, the continuity conditions for the
incoming and outgoing densities $h^{\text{in/out}}(x)$ are
\beq\label{contbefore}
	h^{\text{in}}_j(0)=h^{\text{out}}_j(0),\qquad
    h^{\text{out}}_j(\tfrac{R}{2})=h^{\text{in}}_j(\tfrac{R}{2})
\eeq
and after the connection, these conditions are changed to
\beq\label{contafter}
	h^{\text{in}}_j(0)=h^{\text{out}}_{j+1}(0),\qquad
    h^{\text{out}}_j(\tfrac{R}{2})=h^{\text{in}}_j(\tfrac{R}{2})
\eeq
where here and in the following we understand that leg indices are defined  mod $(N)$.

The former set of continuity conditions holds under time evolution of the
disconnected system, and in averages with the initial state where each system is
independently thermalized. The latter set holds under time evolution of the
connected system, and, as it will turn out, the part of it
at $x=0$ holds in the steady state average (the part of it at $x=R/2$ does not make sense because in the steady state $R$ has been sent to infinity, and we only get averages of local operators around the vertex).

Let us denote by
$H_0^{(j)},\,j=1,\ldots,N$ the Hamiltonians of the disconnected systems on the
legs of the graph. Then the initial density matrix is
\[\rho_0=\frak{n}\lt[e^{-\sum_{j=1}^N\beta_jH_0^{(j)}}\rt]\]
where here and below we use the notation $\frak{n}[\rho] = \rho / \Tr(\rho)$.
This density matrix is invariant under the disconnected-system evolution
Hamiltonian $H_0 = \sum_{j=1}^N H_0^{(j)}$. Assuming that $R\gg x,|t|>0$, the
$H_0$-evolution, taking into account \eqref{contbefore}, is given by
\begin{align}
e^{\ri H_0t}h^{\text{out}}_j(x)
e^{-\ri H_0t}&=\left\{\begin{array}{ll}h^{\text{out}}_j(x-t)& x-t>0\\
h^{\text{in}}_j(t-x)& x-t<0\end{array}\right.\\
e^{\ri H_0t}h^{\text{in}}_j(x)
e^{-\ri H_0t}&=\left\{\begin{array}{ll}h^{\text{in}}_j(x+t)& x+t>0\\
h^{\text{out}}_j(-(x+t))& x+t<0\end{array}\right. .
\end{align}

On the other hand, let us denote by $H$ the Hamiltonian of the connected, total
system. The $H$-evolution, which takes into account the continuity condition
(\ref{contafter}) at the vertex and assuming again that $R\gg x,|t|>0$, is given
by
\begin{align}
e^{\ri Ht}h^{\text{out}}_{j}(x)
e^{-\ri Ht}&=\left\{\begin{array}{ll}h^{\text{out}}_{j}(x-t) &x-t>0\\
h^{\text{in}}_{j-1}(t-x)& x-t<0
\end{array}\right.\\
e^{\ri Ht}h^{\text{in}}_{j}(x)
e^{-\ri Ht}&=\left\{\begin{array}{ll}h^{\text{in}}_j(x+t)& x+t>0\\
h^{\text{out}}_{j+1}(-(x+t))& x+t<0\end{array}\right. .
\end{align}

The average in (\ref{eq:steady_state_limit}) can be written in terms of $H$ and $\rho_0$ as follows:
\[
	\bra\cdots\ket_{R,t_0} = \Tr\left(e^{-\ri Ht_0}\rho_0 e^{\ri Ht_0}
\cdots\right).
\]
Further, the steady state \eqref{eq:steady_state_limit} is invariant under the $H$-evolution, as we will show.

It is clear that in the initial density matrix $\rho_0$, the system decouples into its subsystems
with Hamiltonians $H_0^{(j)},\,j=1,\ldots,N$ on the various legs of the graph.
For instance,
\[
	\bra\prod_{j=1}^N h_j^{\text{in}}(x_j) h_j^{\text{out}}(y_j)\ket_{R,0}
    = \prod_{j=1}^N \bra h_j^{\text{in}}(x_j) h_j^{\text{out}}(y_j)\ket_{R,0}.
\]
In the steady state limit, it turns out that the system again decouples, but not
into the separate legs of the graph. It rather decouples into $N$ subsystems,
for $j=1,\ldots,N$, where in each subsystem $h_j^{\text{in}}$ and
$h_{j+1}^{\text{out}}$ are coupled to each other, in particular satisfying the
first equation of (\ref{contafter}).

\begin{figure}
\centering
\includegraphics[width=0.8\linewidth]{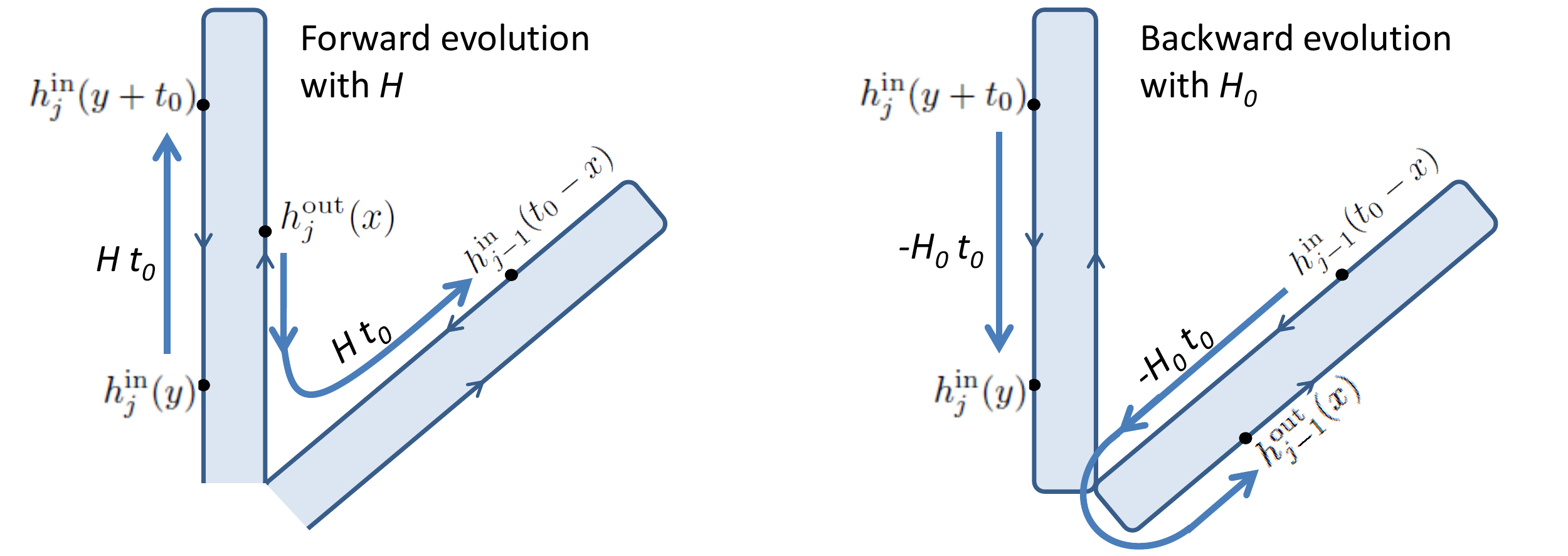}
 \caption{The forward time evolution with $H$ and subsequent backward time evolution with $H_0$ of $h^\text{in}_j(y)$ in equation \eqref{eq:forward_backward_evol_in} and $h^\text{out}_j(x)$ in equation \eqref{eq:forward_backward_evol_out} is shown, focusing only on the $j$-th and $j-1$-th legs of the star graph. This shows that measuring $h^{\text{in}}_j$ and $h^{\text{out}}_j$ in the steady state corresponds to measuring $h^{\text{in}}_j$ and $h^{\text{out}}_{j-1}$ in the disconnected system, respectively.}
 \label{fig:evol}
\end{figure}

Let us now construct explicitly the steady state from
(\ref{eq:steady_state_limit}) using the above considerations, and see explicitly
the $H$-invariance and the decoupling mentioned. This follows the methods of
\cite{bernard2012energy,bernard2013non}. Assuming that $R\gg t_0\gg x>0$, we have
\begin{align}
e^{\ri Ht_0}h^{\text{in}}_j(x)e^{-\ri Ht_0}&=h_j^{\text{in}}(x+t_0)\\
e^{\ri Ht_0}h^{\text{out}}_j(x)e^{-\ri Ht_0}&=h^{\text{in}}_{j-1}(-x+t_0).
\end{align}
Evolving the result backward with $H_0$ then
gives
\begin{align}
      e^{-\ri H_0 t_0}e^{\ri Ht_0} h_j^{\text{in}}(x) e^{-\ri Ht_0}e^{\ri H_0t_0}
    &= h_{j}^{\text{in}}(x) \label{eq:forward_backward_evol_in}\\
	e^{-\ri H_0 t_0}e^{\ri Ht_0} h_j^{\text{out}}(x) e^{-\ri Ht_0}e^{\ri
H_0t_0}
    &= h_{j-1}^{\text{out}}(x). \label{eq:forward_backward_evol_out}
\end{align}
The above proccess of forward evolution with $H$ until the time of connection, and subsequent backward evolution with $H_0$ is visualized in figure \ref{fig:evol} 

For definiteness, let us consider the observable $\prod_{j=1}^N h_j^{\text{in}}(x_j) h_j^{\text{out}}(y_j)$. 
Using the fact that $\rho_0$ is $H_0$-invariant and the above equations, we then obtain
\beq
	\lim_{R\gg t_0\to\infty}
    \bra\prod_{j=1}^N h_j^{\text{in}}(x_j) h_j^{\text{out}}(y_j)\ket_{R,t_0}
    =
    \bra\prod_{j=1}^N
    h_j^{\text{in}}(x_j) h_{j-1}^{\text{out}}(y_j)\ket_{\infty,0}.
    \label{eq:Rt0}
\eeq
This expression is factorized, but the factorization is not in terms of legs. In order to find a steady state that reproduces the above relation, let us introduce the Hamiltonians $H^{(j,j+1)},\,j=1,\ldots,N$, which mutually
commute and which couple together $h_j^{\text{in}}$ with $h_{j+1}^{\text{out}}$
in the same way $H_0^{(j)}$ couple together $h_j^{\text{in}}$ with
$h_{j}^{\text{out}}$. It is convenient for our later derivations to consider
these Hamiltonians, like the Hamiltonians $H_0^{(j)}$, with $R$ finite. This
means that with $H^{(j,j+1)}$ the continuity conditions are
\beq\label{contmix}
	h_j^{\text{in}}(0) = h_{j+1}^{\text{out}}(0),\qquad
    h_{j+1}^{\text{out}}(\tfrac R2) = h_j^{\text{in}}(\tfrac R2),
\eeq
paralleling (\ref{contbefore}). Then we may define
\beq\label{eq:dness}
	\rho_{\text{stat}} = \frak{n}\lt[e^{-\sum_{j=1}^N \beta_j
H^{(j,j+1)}}\rt].
\eeq
We see that any average of an operator $\Or$ with the density matrix $\rho_0$ is equal to the average of a modified operator $\t\Or$ with the density matrix $\rho_{\text{stat}}$, 
where $\t\Or$ is obtained by replacing every $h_j^{\text{out}}$ by $h_{j+1}^{\text{out}}$. Hence we see that the density matrix \eqref{eq:dness} gives rise to the steady state average,
\beq\label{eq:ness}
	\bra\cdots\ket_{\text{stat}} :=
    \lim_{R\to\infty} \Tr\lt(\rho_{\text{stat}}
    \cdots\rt),
\eeq
as (\ref{eq:Rt0}) then implies
\beq
	\lim_{R\gg t_0\to\infty}
    \bra\prod_{j=1}^N h_j^{\text{in}}(x_j) h_j^{\text{out}}(y_j)\ket_{R,t_0}
    =
    \bra \prod_{j=1}^N 
    h_j^{\text{in}}(x_j) h_{j}^{\text{out}}(y_{j})\ket_{\text{stat}}
\eeq
A similar derivation holds for any other finite product of fields.
Since the Hamiltonians $H^{(j,j+1)}$ commute, this shows the factorization mentioned above. The result below of
a nonzero energy current then confirms that this is a non-equilibrium steady state.

\begin{figure}[h!]
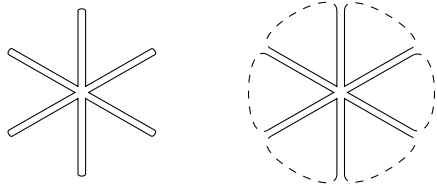

\centering
\tikz[anchor=base,baseline]{
	\draw[thin]
    	(0.05,1.08)--(0.05,0.08)--(0.916,0.58)
        (-0.05,1.08)--(-0.05,0.08)--(-0.916,0.58)
        (0.96,0.5)--(0.09,0)--(0.96,-0.5)
        (-0.96,0.5)--(-0.09,0)--(-0.96,-0.5)
        (0.05,-1.08)--(0.05,-0.08)--(0.916,-0.58)
        (-0.05,-1.08)--(-0.05,-0.08)--(-0.916,-0.58)
		(0.05,1.08)to[in=90,out=90](-0.05,1.08)
        (0.916,0.58)to[in=20,out=20](0.96,0.5)
        (-0.916,0.58)to[in=150,out=150](-0.96,0.5)
        (-0.96,-0.5)to[in=-150,out=-150](-0.916,-0.58)
        (0.05,-1.08)to[in=-90,out=-90](-0.05,-1.08)
        (0.96,-0.5)to[in=-20,out=-20](0.916,-0.58);
}
$\qquad$
\tikz[anchor=base,baseline]{
	\draw[thin]
    	(0.05,1.08)--(0.05,0.08)--(0.916,0.58)
        (-0.05,1.08)--(-0.05,0.08)--(-0.916,0.58)
        (0.96,0.5)--(0.09,0)--(0.96,-0.5)
        (-0.96,0.5)--(-0.09,0)--(-0.96,-0.5)
        (0.05,-1.08)--(0.05,-0.08)--(0.916,-0.58)
        (-0.05,-1.08)--(-0.05,-0.08)--(-0.916,-0.58);
	\draw[thin,dashed]
		(0.05,1.08)to[in=20,out=90](0.916,0.58)
        (-0.05,1.08)to[in=150,out=90](-0.916,0.58)
        (0.96,0.5)to[in=-20,out=20](0.96,-0.5)
        (-0.96,0.5)to[in=-150,out=150](-0.96,-0.5)
        (0.05,-1.08)to[in=-20,out=-90](0.916,-0.58)
        (-0.05,-1.08)to[in=-150,out=-90](-0.916,-0.58);
}
\caption{On the left, the physical situation at finite $R$ after the connection: several heat baths
connected at a point. The Hamiltonian $H$ represents evolution along the path
going around the graph. On the right, how expectation values are calculated
using the fact that in the steady state limit the system decouples into
subsystems as in this picture. The Hamiltonians $H^{(j,j+1)}$ represent
evolutions along the distinct paths.
These two time-evolutions are the same for fields near the vertex.}
\label{fig:connect}
\end{figure}

A picture representing, side by side, the physical situation after connection,
and the Hamiltonians $H^{(j,j+1)}$ used in the construction of the steady-state
density matrix, is shown in Figure \ref{fig:connect}.

In the $H^{(j,j+1)}$ subsystems, it is simpler to
work with ``right-moving'' fields only, which are the incoming fields from one
leg and the outgoing fields into the next leg. We will label these by the two
legs on which they move, and make the following identification:
\begin{equation}\label{eq:h_chiral}
h^{(j,j+1)}(x):=\left\{\begin{array}{ll}h^{\text{in}}_j(-x)& x<0\\
h_{j+1}^{\text{out}}(x)& x\geq0\end{array}\right.,\qquad
x\in\left[-\tfrac{R}{2},\tfrac{R}{2}\right].
\end{equation}
With these new fields, the time evolution is now simply
\begin{equation}
e^{\ri Ht}h^{(j,j+1)}(x)e^{-\ri Ht}=h^{(j,j+1)}(x-t), \qquad x,t\ll R
\label{eq:time_evolution}
\end{equation}
and $h^{(j,j+1)}(x)$ is continuous.

\begin{rema}
 It is important to note that the connected-system Hamiltonian $H$ {\em does not}
commute with $H^{(j,j+1)}$. However, thanks to the agreement between the first
equations of (\ref{contafter}) and (\ref{contmix}), the evolution with $H$ by a
time $t$ on fields at $x$, for any $R\gg |t|,x>0$, is exactly in agreement with
the evolution with $\sum_{j=1}^N H^{(j,j+1)}$:
\[
	H = \sum_{j=1}^N H^{(j,j+1)} \quad\mbox{as evolution operators on fields a finite distance from the vertex}.
\]
Since $\sum_{j=1}^N H^{(j,j+1)}$ commutes with $\rho_{\text{stat}}$, this
implies that the steady state average $\bra\cdots\ket_{\text{stat}}$ (defined only for products of local fields a finite distance from the vertex) is
$H$-invariant, as claimed above.
\end{rema}


\section{Results and discussion} \label{sect:2}

We now state our main results concerning the energy current and its fluctuations
in the steady state described in the previous section.

Consider a charge
\[
	Q=\sum_{j=1}^N\alpha_jH^{(j)}_0,
\]
the weighted sum of the
energies in the various legs of the star graph. The time derivative of $Q$ is
the associated ``energy current'' operator, ${\cal J}:=i[H,Q] = \sum_{j=1}^N
\alpha_j (h^{\text{out}}_j(0) - h^{\text{in}}_j(0))$, which is the weighted sum
of the momentum densities on the various legs of the graph. It is local, hence
it has a well-defined steady-state average $J = \bra{\cal J}\ket_{\text{stat}} =
i\bra [H,Q]\ket_{\text{stat}}$. Note that this steady-state average does not
necessarily vanish because $Q$ itself is not a local operator. 

\subsection{Average current}

The average current is in fact simple to evaluate using standard CFT techniques
(see for instance \cite{CFTbook}) which give the result $\bra h^{(j,j+1)}\ket_{\text{stat}} = \pi
c/(12\beta_j^2)$. Hence:
\beq\label{J}
	J=\frac{\pi c}{12}\sum_{j=1}^N\frac{\Delta\alpha_j}{\beta_j^2}
\eeq
where $\Delta\alpha_j:=\alpha_{j+1}-\alpha_j$. This generalizes the result of
\cite{bernard2012energy}, where the case $N=2$ and $\alpha_1=-\alpha_2=1/2$ was
considered. This is in agreement with the simple picture according to which
energy flows from leg $j$ to leg $j+1$ with the information of the asymptotic
thermal bath of leg $j$, with a current $\pi c/(12 \beta_j^2)$.

The charge and energy current has been calculated via a scattering state
approach in \cite{mintchev2011non,mintchev2013luttinger} in the context of free fermion and Luttinger liquid models for
more general continuity conditions at the vertex, where the systems also have a
set of chemical potentials. As an example, for a system of Dirac fermions (with
central charge $c=1$) the energy current from one leg into the vertex (hence
into all other legs) was calculated. Our result agrees with the result found in
\cite{mintchev2011non,mintchev2013luttinger} for a critical system with our sequential boundary
conditions and all the chemical potentials set to zero. For more details, see
Appendix A.

\subsection{Current fluctuations}

The energy current fluctuations can also be evaluated within our framework.
There are various ways of defining these fluctuations. We will consider the
setup \cite{esposito2009nonequilibrium} where the charge $Q$ is first
quantum-mechanically measured (von Neumann measurement) at the contact time
$-t_0$, and then measured at time 0. The cumulants of the difference
$\Delta_{t_0} Q$ of the measured values are then evaluated in the steady state
limit (where in particular $t_0$ becomes infinite). We find that all cumulants
diverge linearly in $t_0$, and we obtain the exact coefficients of this
divergence for all cumulants. These can be organized into the coefficients of
the Taylor expansion in $i\lambda$ of the scaled cumulant-generating function
$F(\lambda)$, which is the Legendre transform of the large deviation function. As was shown in \cite{bernard2013non} (this was the case for $N=2$, but the proof is
easily generalizable to all $N$), the result of these
operations can be represented 
by the formula
\beq\label{Fla}
	F(\lambda) = \lim_{t\to\infty} \frc1t
	\log \bra e^{\ri \lambda Q(t)} e^{-\ri \lambda Q} \ket_{
	\text{stat}}
\eeq
where $Q(t) = e^{\ri Ht} Q e^{-\ri Ht}$. This is of the same form as the standard
expression for the so-called full-counting statistics of charge transfer, which was
first obtained within the context of indirect measurements, instead of the
two-time von Neuman measurement protocol we discuss above, 
in free fermion models in \cite{LL}. In fact, we will show that formula
\eqref{Fla} is equivalent to the simpler, ``naive'' expression
\beq\label{Fnaive}
	F(\lambda) = \lim_{t\to\infty} \frc1t
	\log \bra e^{\ri \lambda (Q(t)-Q)}  \ket_{
	\text{stat}}.
\eeq
This, for instance, immediately implies that the second order in $\ri \lambda$, which is the noise (made up of the thermal noise and the
so-called shot noise), takes the familiar form:
\beq\label{fqtq}
	F(\lambda) = \ri \lambda J + \frc{(\ri \lambda)^2}2
	\int_0^\infty dt\,\lt(
	\bra\{{\cal J}(t),{\cal J}(0)\}\ket_{\text{stat}} -
	2J^2 \rt)+\ldots\,.
\eeq
Further, using the time evolution equations of the previous section, one finds
that
\beq\label{qtq}
	\Delta_t Q := Q(t) - Q = \sum_j \Delta \alpha_j \int_0^t dx\,
h^{\text{in}}_j(x).
\eeq
Hence, the function $F(\lambda)$ can also be interpreted as the generating
function for large-$L$ cumulants of the weighted sum, over the legs of the
graph, of the incoming energies $\int_0^L dx\, h^{\text{in}}_j(x)$ on
intervals of lengths $L$. In contrast with the initial two-measurement
description, in this interpretation, these incoming energies are now measured in
single von Neumann measurements.

We find the following exact expression for $F(\lambda)$:
\begin{equation}\label{cgf}
F(\lambda)=\frac{\ri \lambda\pi
c}{12}\sum_{j=1}^N\frac{\Delta\alpha_j}{\beta_j(\beta_j-\ri
\Delta\alpha_j\lambda)},
\end{equation}
generalizing the result of \cite{bernard2012energy} to higher $N$ and to
different charges $Q$ (different weighted sums). Specializing to the case where the charge measured is the energy difference
between two contiguous parts of the star graph $Q=\frac{1}{2}(H_{\text{part
1}}-H_{\text{part 2}})$, the cumulant generating function simplifies to:
\begin{equation}\label{Fpart}
F(\lambda)=\frac{\ri \lambda\pi c}{12}\left(\frac{1}{\beta_m(\beta_m-\ri
\lambda)}-\frac{1}{\beta_n(\beta_n+\ri \lambda)}\right)
\end{equation}
where $m$ and $n$ are the legs just after which the sign of the weight changes
from minus to plus and from plus to minus, respectively. This result is
identical in form to the case $N=2$.

\subsection{Fluctuation relations and Poisson processes}

The function \eqref{Fpart} satisfies the usual symmetry relation
(or fluctuation relation) found in bipartite systems,
\begin{equation}\label{fr}
F(\ri (\beta_n-\beta_m)-\lambda)=F(\lambda).
\end{equation}
This is a standard relation in the context of non-equilibrium steady states, and
is characteristic of an exponentially decaying ratio of probabilities
for energy transfer from the lower-temperature (larger $\beta$) region and
energy transfer from the higher-temperature (smaller $\beta$) region
\cite{esposito2009nonequilibrium}, over large periods of time. That is, let $P_{\text{(2 parts)}}(q)$ be the
probability that $\Delta_{t_0} Q = q$ for this particular choice of weights
$\alpha_j$. Then \eqref{fr} is equivalent to a so-called \emph{steady-state} fluctuation theorem:
\begin{equation}
P_{\text{(2 parts)}}(q)\sim e^{(\beta_n-\beta_m)q}P_{\text{(2 parts)}}(-q)
\label{eq:fluctrel}
\end{equation}
where ``$\sim$'' indicates that the fluctuation relation holds only at large $t_0$ and accordingly large $q$.

In general, however, the cumulant generating function \eqref{cgf} does not
satisfy such a simple symmetry relation involving a shift in $\lambda$ only. Yet
there is a simple physical picture behind formula \eqref{cgf}, and a generalized
symmetry relation that agrees with this picture. Denoting $\omega_j :=
\lambda\Delta\alpha_j$, with $j=1,\ldots,N$, the symmetry relation is:
\beq\label{gensym}
	F(\omega_1-\ri \beta_1,\ldots,\omega_N-\ri\beta_N)
	\quad\mbox{is symmetric under permutations of 
	$\omega_1,\ldots,\omega_N$}.
\eeq
Indeed, we observe that our exact formula \eqref{cgf} satisfies \eqref{gensym}:
\beq\label{eqgs}
	F(\omega_1-\ri\beta_1,\ldots,\omega_N-\ri\beta_N) =
	-\frc{\pi c}{12} \sum_j \lt(\frc1{\beta_j} + \frc1{\ri\omega_j}\rt).
\eeq

The physical picture is as follows.
Due to the separation of the stationary state into the $H^{(j,j+1)}$
subsystems, we expect $F(\lambda)$ in general to correspond to a set of $N$
independent processes whereby energy is transferred from leg $j$ to leg $j+1$,
with $j=1,\ldots,N$. Let $P^{(j,j+1)}(r)$ be the probability that an energy
$r>0$ be transferred from leg $j$ to leg $j+1$ in such a process (this energy is
always positive thanks to our choice of continuity conditions at the vertex). Further, let
us  denote by $P_{\Delta \alpha_1,\ldots,\Delta \alpha_N}(q)$ the probability
that $\Delta_{t_0}Q = q$ for general weights $\alpha_j$ (it only depends on the
weight differences $\Delta\alpha_j$). We have in particular $e^{t_0 F(\lambda)}
= \int dq\,e^{\ri \lambda q}P_{\Delta \alpha_1,\ldots,\Delta \alpha_N}(q)$.
Then we expect that at large $t_0$, we have
\beq
	P_{\Delta \alpha_1,\ldots,\Delta \alpha_N}(q)
	\sim \int_{q = \sum_j \Delta \alpha_j r_j}
	\prod_j dr_j\,P^{(j,j+1)}(r_j).
\eeq
Furthermore, we expect the independent $j\to j+1$ processes to be related to each
other via a similar fluctuation relation as \eqref{eq:fluctrel}:
\beq\label{gensym1}
	e^{\beta_j r} P^{(j,j+1)}(r) \sim e^{\beta_k r}
P^{(k,k+1)}(r)\quad\forall
	\quad j,k.
\eeq

We can verify that in the 2-part case discussed above, we do recover
\eqref{eq:fluctrel}:
\beqa
	e^{\beta_m q} P_{\text{(2 parts)}}(q)&\sim&
	\int_{q = r_m-r_n} dr_m dr_n\,
	e^{\beta_m (r_m-r_n)} P^{(m,m+1)}(r_m) P^{(n,n+1)}(r_n) \n
	&\sim& \int_{q = r_m-r_n} dr_m dr_n\,
	e^{\beta_n (r_m-r_n)} P^{(n,n+1)}(r_m) P^{(m,m+1)}(r_n) \n
	&=& \int_{-q = r_m-r_n} dr_m dr_n\,
	e^{\beta_n (r_n-r_m)} P^{(n,n+1)}(r_n) P^{(m,m+1)}(r_m) \n
	&\sim& e^{\beta_n q} P_{\text{(2 parts)}}(-q).\no
\eeqa
What's more, this picture implies the generalized symmetry relation \eqref{gensym} for
$F(\lambda)$.
This is simple to see from
\beq
	e^{t_0 F(\omega_1,\ldots,\omega_N)}
	\sim \int \prod_j dr_j\, e^{\ri\omega_j r_j} P^{(j,j+1)}(r_j)
\eeq
and using (\ref{gensym1}).

Hence, our result for the large-time cumulant generating function is indeed in
agreement with the proposed picture.

Consider the conditions that (i) $F(\lambda)$ separates into a sum over $j$ of a
two-variable function of $\omega_j:=\lambda \Delta\alpha_j$ and $\beta_j$; (ii)
$F(\lambda)$ is a homogeneous function of $\omega_j$ and $\beta_j$ of degree -1;
(iii) $F(0) = 0$; and (iv) the symmetry relation (\ref{gensym}) holds. We
observe that these conditions are sufficient to fully fix the function
$F(\lambda)$ to the form (\ref{cgf}), up to an overall normalization. Indeed,
the first condition says that $F(\lambda) = \sum_j f(\omega_j,\beta_j)$. Let us
denote $\t f(\omega,\beta) = f(\omega-\ri\beta,\beta)$. Then the fourth
condition implies that $\t f(\omega,\beta) = \t f_1(\omega) + \t f_2(\beta)$. By
the second condition, we then have $\t f_1(\omega) = a_1/(\ri\omega)$ and $\t
f_2(\beta) = a_2/\beta$. Further, the third condition says that $\t
f(\ri\beta,\beta)=0$, which gives $a_1=a_2$. Hence we indeed find the right-hand
side of (\ref{eqgs}) for $\sum_j \t f_j(\omega_j,\beta_j)$ up to a
normalization, which completes 
the 
proof. Condition (i) is a consequence of the factorization of the stationary
density matrix \eqref{eq:dness} and of (\ref{qtq}) and (\ref{fqtq}). Condition
(ii) is a consequence of scale invariance, and condition (iii) of the basic
definition of the generating function. Hence, along with basic properties of
CFT, the symmetry relation (\ref{gensym}) fully fixes the cumulant generating
function $F(\lambda)$. This generalizes what was observed in \cite{bernard2012energy} in the case
$N=2$.

We finally note that the independent processes of energy transfer from legs $j$
and $j+1$, with probabilities $P^{(j,j+1)}(q)$, in the asymptotic regime
$t_0\to\infty$, can be uniquely identified. Indeed, we can interpret
$F(\lambda)$ as describing the cumulants of a random variable $q$ coming from
classical processes as follows: at each interface $j\to j+1$, there is a family
of independent Poisson processes parametrized by $r>0$, with intensities
$e^{-\beta_j r}$, each contributing to the random variable $q$ a value $\Delta
\alpha_j r$. This is again a generalization of what was observed in \cite{bernard2012energy} in
the case $N=2$, and further confirms that for large energy transfers, CFT is
equivalently described by replacing right-movers and left-movers by independent
carriers of energy units, distributed according to the appropriate density of
state, jumping towards the right and left, respectively, in a Poissonian
fashion.


\section{The continuous Virasoro algebra and its diagrams} \label{sect:3}

Our strategy to prove the form (\ref{cgf}) of the cumulant generating function
will be an analysis of \eqref{Fla} as a series expansion in $\ri \lambda$, along
with the use of the Virasoro algebra at the basis of CFT. This algebraic
analysis provides an alternative to the local-variable analysis used in \cite{bernard2013non} in order to
prove (\ref{cgf}) in the case $N=2$ (and for $\alpha_1=-\alpha_2=1/2$). The
Virasoro algebra occurs as a natural algebra in the quantization of CFT on the
circle or on segments (or in radial quantization). Of course, in the
construction discussed in Section \ref{sect:1}, we have instead, after the
connection, the star-graph geometry; yet the usual notions of CFT on segments of
lengths $R/2$ can still be used.

A complete analysis of the quench using the standard formulation of the Virasoro algebra would be interesting.
Some initial ideas are presented in Appendix \ref{sect:appB}, where it turns out we obtain the 
expected value of the holomorphic dimension of the branch point twist field \cite{CC},
which is used in entanglement entropy.

However, in the establishment of the steady state through the steady-state limit
\eqref{eq:steady_state_limit}, the large-$R$ limit needs to be taken before the
time evolution can be analyzed. The large-$R$ limit of the quantization on the
circle or on segments gives rise to the quantization on the line or the semiline
(depending on how exactly the limit is taken). In this quantization space, the
Virasoro algebra is no longer a natural algebra. At equilibrium, one can
circumvent this problem, as one knows the Euclidean geometry describing the
large-$R$ limit: either a cylinder, infinite or semi-infinite (at finite
temperature, with the circumference equal to the inverse temperature), or the
plane or half-plane (at zero temperature). In these cases, either using the
quantization on the circle or radial quantization, one can still make use of the
Virasoro algebra. However, out of equilibrium, there is not yet a clear
Euclidean geometric picture; hence one needs to keep the quantization on the
line or the semi-
line (here, appropriately tailored to the graph geometry), whence one loses the
Virasoro algebra.

The natural algebraic structure that emerges in the large-$R$ limit is that
which one may refer to as the {\em continuous Virasoro algebra}. This is an
algebra similar to the Virasoro algebra, but with a continuous index. The
algebraic structure is rather simple, but its connection to the result of the
large-$R$ limit of the Virasoro algebra present some subtleties. Its
representation theory also presents many subtleties, which we will not address
here. We present instead basic aspects of this algebra and the diagrams used to
evaluate certain traces of product of algebra elements, which we will need in
the next section.

\subsection{Definitions}
Consider a Lie algebra with a continuous basis
\begin{equation}
\{a_p:p\in\mathbb{R}\}
\label{eq:continuum_algebra}
\end{equation}
satisfying the following commutation relations
\begin{equation}\label{contalg}
[a_p,a_q]=(p-q)a_{p+q}+(2p\,k+\tfrac{c}{12}p^3)\delta(p+q)
\end{equation}
where\footnote{Instead of seeing $k,c$ as numbers, we could also see them as
additional central elements; but in the representation we will need these simply
take fixed values.} $k,c\in\R$. For a fixed $c$, the algebras corresponding to
different values of $k$ are isomorphic: a simple change of basis $\t a_p= a_p +
k\delta(p)$ makes the linear-in-$p$ term vanish. However it is convenient for us
to keep this term. Given an appropriate highest-weight representation of this
algebra, where $a_0$ is diagonalizable and with eigenvalues that are bounded
from below, we may define a state by the following ratio of traces
\begin{equation}
\langle\cdots\rangle_\beta\equiv\frac{\text{Tr}\left(e^{-2\pi\beta
a_0}\cdots\right)}{\text{Tr}\left(e^{-2\pi\beta a_0}\right)}
\end{equation}
where $\cdots$ represents some product of the generators $a_{p}$, and $\beta>0$
is some parameter. A simple calculation using \eqref{contalg}, and assuming that
the cyclic property of the trace holds,
\[\langle a_p\rangle_\beta=\frac{\text{Tr}\left(e^{-2\pi\beta
a_0}a_p\right)}{\text{Tr}\left(e^{-2\pi\beta
a_0}\right)}=\frac{\text{Tr}\left(a_pe^{-2\pi\beta
a_0}\right)}{\text{Tr}\left(e^{-2\pi\beta a_0}\right)}=e^{-2\pi\beta p}\langle
a_p\rangle_\beta=0\qquad\text{for }p\neq 0,\]
shows that we must have
\beq
	\bra a_p \ket_\beta = B \,\delta(p)
\eeq
for some number $B$. With an appropriate choice of basis, it is possible to
impose
\[
	B=0.
\]
With this condition, the basis is completely fixed (albeit in a state-dependent
way) and in particular $k$ in (\ref{contalg}) is unambiguous. Below we will have
a representation where with the condition $B=0$,
\begin{equation}\label{kbeta}
 k = k(\beta):=\frac{c}{24\beta^2}.
\end{equation}

We can then calculate expectation values of products of more than one generator,
again using the cylic property of the trace along with the algebra relations.
For example,
\begin{multline}\langle
a_{p_1}a_{p_2}\rangle_{\beta}=\frac{\text{Tr}\left(e^{-2\pi\beta
a_0}a_{p_1}a_{p_2}\right)}{\text{Tr}\left(e^{-2\pi\beta
a_0}\right)}=\frac{\text{Tr}\left(a_{p_2}e^{-2\pi\beta
a_0}a_{p_1}\right)}{\text{Tr}\left(e^{-2\pi\beta a_0}\right)}=e^{-2\pi\beta
p_2}\langle a_{p_2}a_{p_1}\rangle_{\beta}=\frac{1}{e^{2\pi\beta
p_2}-1}\langle[a_{p_2},a_{p_1}]\rangle_\beta\\
=\frac{1}{e^{2\pi\beta
p_2}-1}\left(2p_2\,k(\beta)+\frac{c}{12}p_2^3\right)\delta(p_1+p_2).
\end{multline}

\subsection{Diagrams}

Expectation values of products of several generators $a_p$ can be calculated
similarly using the cyclic property of traces on, for instance, the rightmost
generator, giving
\begin{equation}
\langle a_{p_1}\cdots a_{p_M}\rangle_\beta=\frac{1}{e^{2\pi\beta
p_M}-1}\sum_{j=1}^{M-1}\langle a_{p_1}\cdots[a_{p_M},a_{p_j}]\cdots
a_{p_{M-1}}\rangle_\beta.
\label{eq:iteration}
\end{equation}
One obtains a recursion relation for these expectation values by using
\eqref{contalg} for the commutator. There are two terms occurring: the first is
proportional to the generator $a_{p_j+p_M}$, and the second is the central term,
with a delta function factor $\delta(p_j+p_M)$. We may represent these two
contributions using the following diagrams:
\begin{subequations}
\begin{align}
\tikz[scale=1.5,anchor=base,baseline=0]{
\insertion{0}{0}{1}{.05}{.75}
\insertion{0}{0}{2}{.05}{.75}
\vertnonzero{0}{0}{1}{1}{2}
}
&\qquad=\quad\frac{p_2-p_1}{e^{2\pi\beta p_2}-1}\quad
\label{eq:diagram_rule_conn}\\
\tikz[scale=1.5,anchor=base,baseline]{
\vertzero{0}{0}{1}{1}{2}
\insertion{0}{0}{1}{.05}{.75}
\insertion{0}{0}{2}{.05}{.75}
}
&\qquad=\quad\frac{1}{e^{2\pi\beta
p_2}-1}\left(2p_2\,k(\beta)+\tfrac{c}{12}p_2^3\right)\delta(p_1+p_2).
\label{eq:diagram_rule_disconn}
\end{align}
\end{subequations}
Repeating the process, we obtain the expectation value as a sum of diagrams
constructed according to the following rules:
\begin{enumerate}
\item Start with a horizontal alignment of ``open dots'' carrying momenta
$p_1,\ldots,p_M$, for instance
\[
\tikz[scale=1.5,anchor=base,baseline=0]{
\insertion{0}{0}{1}{.05}{.75}
\insertion{0}{0}{2}{.05}{.75}
\insertion{0}{0}{3}{.05}{.75}
\insertion{0}{0}{4}{.05}{.75}
\insertion{0}{0}{5}{.05}{.75}
}
\]
\item Connect the rightmost dot with a dot to its left, using either the vertex
\eqref{eq:diagram_rule_conn} or the cap \eqref{eq:diagram_rule_disconn}. The use
of the vertex \eqref{eq:diagram_rule_conn} leaves an open dot at the horizontal
position of the leftmost dot in the connected pair, and at one vertical step
higher, carrying the sum of momenta of the connected dots. The use of the cap
\eqref{eq:diagram_rule_disconn} closes both dots connected. If there are only
two open dots, use only \eqref{eq:diagram_rule_disconn}.
\item Repeat the previous step with the remaining open dots, one vertical step
higher.
\item Finish when there are no remaining open dots.
\end{enumerate}
Note that we do not allow for diagrams with a single open dot; this corresponds
to our choice of basis according to which $\langle a_p\rangle=0$. 
As an example, two diagrams contributing to the expectation value $\langle
a_{p_1}a_{p_2}a_{p_3}a_{p_4}a_{p_5}\rangle_\beta$ are
\[
\tikz[scale=1.4,anchor=base,baseline=0]{
        \insertion{0}{0}{1}{.05}{.75}
        \insertion{0}{0}{2}{.05}{.75}
        \insertion{0}{0}{3}{.05}{.75}
        \insertion{0}{0}{4}{.05}{.75}
        \insertion{0}{0}{5}{.05}{.75}
        \vertzero{0}{0}{4}{1}{2}
		\draw[thin, rounded
corners=2pt](0+2,0+0.2*3)--(0+3,0+0.2*3)--(0+3,0);\draw[thin](0+2,0)--(0+2,
0+0.15+0.2*3);\draw[fill=gray](0+2,0+0.2*3) circle
(0.05);\fill[black](0+2+.06,0+.1+0.2*3) node[anchor=west,scale=.75] 
{$p_{2}+p_{3}+p_{4}+p_{5}$};
		\vertnonzero{0}{0}{2}{3}{4}
		\vertnonzero{0}{0}{1}{2}{5}
}\qquad\text{and}\qquad
\tikz[scale=1.4,anchor=base,baseline=0]{
		\insertion{0}{0}{1}{.05}{.75}
        \insertion{0}{0}{2}{.05}{.75}
        \insertion{0}{0}{3}{.05}{.75}
        \insertion{0}{0}{4}{.05}{.75}
        \insertion{0}{0}{5}{.05}{.75}
		\vertzero{0}{0}{3}{1}{3}
        \vertzero{0}{0}{2}{2}{4}
        \vertnonzero{0}{0}{1}{3}{5}\;.
}
\]
The first diagram corresponds to the term
\begin{multline} \label{eq:connected_value}
\frac{p_5-p_2}{e^{2\pi\beta p_5}-1}\frac{p_4-p_3}{e^{2\pi\beta
p_4}-1}\frac{p_3+p_4-(p_2+p_5)}{e^{2\pi\beta(p_3+p_4)}-1}\\ 
\times\frac{2(p_2+p_3+p_4+p_5)k(\beta)+\tfrac{c}{12}(p_2+p_3+p_4+p_5)^3}{e^{
2\pi\beta(p_2+p_3+p_4+p_5)}-1}\delta(p_1+p_2+p_3+p_4+p_5),
\end{multline}
and the second diagram corresponds to the term
\beq \label{eq:disconnected_value}
	\frac{p_5-p_3}{e^{2\pi\beta
p_5}-1}\frac{2(p_3+p_5)k(\beta)+\tfrac{c}{12}(p_3+p_5)^3}{e^{2\pi\beta(p_3+p_5)}
-1}\delta(p_1+p_3+p_5)\frac{2p_4\,k(\beta)+\tfrac{c}{12}p_4^3}{e^{2\pi\beta
p_4}-1}\delta(p_2+p_4).
\eeq
Note that the second diagram can be written as a product of two separate
diagrams
\[
\tikz[scale=1.5,anchor=base,baseline=0]{
	\insertion{0}{0}{1}{.05}{.75}
    \insertion{2}{0}{2}{.05}{.75}
    \insertion{-1}{0}{3}{.05}{.75}
    \insertion{1}{0}{4}{.05}{.75}
    \insertion{-2}{0}{5}{.05}{.75}
	\vertzero{0}{0}{1}{4}{5}
    \vertzero{0}{0}{2}{1}{2}
	\draw[thin, rounded corners=2pt]
    	(2,0.2*1)--(3,0.2*1)--(3,0);
    \draw[thin]
    	(2,0)--(2,0.15+0.2*1);
    \draw[fill=gray]
    	(2,0.2*1) circle (0.05);
    \fill[black]
    	(2+.06,0.1+0.2*1) node[anchor=west,scale=.75] {$p_{3}+p_{5}$};
}
\]

We will call a diagram ``connected" if there is a connected path between all
initial dots; otherwise it is ``disconnected''. Above, the first diagram is
connected, while the second is disconnected. It is clear, from our diagrammatic
rules, that a disconnected diagram can always be written as a product of
connected diagrams. Note also that the value of the momentum is conserved at
each three-leg vertex introduced by the first type of connection
\eqref{eq:diagram_rule_conn}, and that the second type of connection
\eqref{eq:diagram_rule_disconn} produces a factor of a delta function that sets
the sum of the momenta to zero. Hence, the set of momenta in every connected
diagram is constrained to sum to zero, but has no other constraint.

As another example, the full expression for the expectation value $\bra a_{p_1}
a_{p_2} a_{p_3} a_{p_4}\ket_\beta$ is
\begin{multline}
\tikz[scale=1.4,anchor=base,baseline=0]{
	\insertion{0}{0}{1}{.05}{.75}
    \insertion{0}{0}{2}{.05}{.75}
    \insertion{0}{0}{3}{.05}{.75}
    \insertion{0}{0}{4}{.05}{.75}
	\vertzero{0}{0}{3}{1}{2}
    \draw[thin, rounded corners=2pt](1,0.2*2)--(3,0.2*2)--(3,0);\draw[thin](1,0)--(1,0.15+0.2*2);
    \draw[fill=gray](1,0.2*2) circle(0.05);\fill[black](1+.06,.1+0.2*2) node[anchor=west,scale=.75] {$p_{1}+p_{3}+p_{4}$};
    \vertnonzero{0}{0}{1}{1}{4}
}+
\tikz[scale=1.4,anchor=base,baseline=0]{
	\insertion{0}{0}{1}{.05}{.75}
    \insertion{0}{0}{2}{.05}{.75}
    \insertion{0}{0}{3}{.05}{.75}
    \insertion{0}{0}{4}{.05}{.75}
	\vertzero{0}{0}{3}{1}{2}
    \vertnonzero{0}{0}{2}{2}{3}
    \vertnonzero{0}{0}{1}{1}{4}
}+
\tikz[scale=1.4,anchor=base,baseline=0]{
	\insertion{0}{0}{1}{.05}{.75}
    \insertion{0}{0}{2}{.05}{.75}
    \insertion{0}{0}{3}{.05}{.75}
    \insertion{0}{0}{4}{.05}{.75}
	\vertzero{0}{0}{3}{1}{2}
    \vertnonzero{0}{0}{2}{1}{3}
    \vertnonzero{0}{0}{1}{2}{4}
}+\\
\tikz[scale=1.4,anchor=base,baseline=0]{
	\insertion{0}{0}{1}{.05}{.75}
    \insertion{0}{0}{2}{.05}{.75}
    \insertion{0}{0}{3}{.05}{.75}
    \insertion{0}{0}{4}{.05}{.75}
	\vertzero{0}{0}{3}{1}{2}
    \draw[thin, rounded corners=2pt](2,0.2*2)--(3,0.2*2)--(3,0);\draw[thin](2,0)--(2,0.15+0.2*2);
    \draw[fill=gray](2,0.2*2) circle(0.05);\fill[black](2+.06,.1+0.2*2) node[anchor=west,scale=.75]{$p_{2}+p_{3}+p_{4}$};
    \vertnonzero{0}{0}{1}{2}{4}
}+
\tikz[scale=1.4,anchor=base,baseline=0]{
	\insertion{0}{0}{1}{.05}{.75}
    \insertion{0}{0}{2}{.05}{.75}
    \insertion{0}{0}{3}{.05}{.75}
    \insertion{0}{0}{4}{.05}{.75}
	\vertzero{0}{0}{3}{1}{2}
    \draw[thin, rounded corners=2pt](1,0.2*2)--(3,0.2*2)--(3,0);\draw[thin](1,0)--(1,0.15+0.2*2);
    \draw[fill=gray](1,0.2*2) circle(0.05);\fill[black](1+.06,.1+0.2*2) node[anchor=west,scale=.75]{$p_{1}+p_{2}+p_{3}+p_{4}$};
    \vertnonzero{0}{0}{1}{3}{4}
}+
\tikz[scale=1.4,anchor=base,baseline=0]{
	\insertion{0}{0}{1}{.05}{.75}
    \insertion{0}{0}{2}{.05}{.75}
    \insertion{0}{0}{3}{.05}{.75}
    \insertion{0}{0}{4}{.05}{.75}
	\vertzero{0}{0}{3}{1}{2}
    \draw[thin, rounded corners=2pt](2,0.2*2)--(3,0.2*2)--(3,0);\draw[thin](2,0)--(2,0.15+0.2*2);
    \draw[fill=gray](2,0.2*2) circle(0.05);\fill[black](2+.06,.1+0.2*2) node[anchor=west,scale=.75]{$p_{2}+p_{3}+p_{4}$};
    \vertnonzero{0}{0}{1}{3}{4}
}+\\
\tikz[scale=1.4,anchor=base,baseline=0]{
	\insertion{0}{0}{1}{.05}{.75}
    \insertion{0}{0}{2}{.05}{.75}
    \insertion{0}{0}{3}{.05}{.75}
    \insertion{0}{0}{4}{.05}{.75}
	\vertzero{0}{0}{2}{2}{3}
    \vertzero{0}{0}{1}{1}{4}
}+
\tikz[scale=1.4,anchor=base,baseline=0]{
	\insertion{0}{0}{1}{.05}{.75}
    \insertion{0}{0}{2}{.05}{.75}
    \insertion{0}{0}{3}{.05}{.75}
    \insertion{0}{0}{4}{.05}{.75}
	\vertzero{0}{0}{2}{1}{3}
    \vertzero{0}{0}{1}{2}{4}
}+
\tikz[scale=1.4,anchor=base,baseline=0]{
	\insertion{0}{0}{1}{.05}{.75}
    \insertion{0}{0}{2}{.05}{.75}
    \insertion{0}{0}{3}{.05}{.75}
    \insertion{0}{0}{4}{.05}{.75}
	\vertzero{0}{0}{2}{1}{2}
    \vertzero{0}{0}{1}{3}{4}
}\label{example_diagram}
\end{multline}

We finally note that diagrams obtained according the the above rules have a nice
dynamic interpretation. Indeed, we imagine starting with $N$ particles, carrying
conserved quantities $p_j$, which can only move left, such that two particles
can either jump through each other without interacting, or interact with each
other by forming a bound state (a new particle) or by annihilating each other.
Observing diagrams from bottom to top and interpreting them as world-lines, with
time increasing upwards, the rules above give rise to all possible inequivalent
events for this dynamics with the further constraint that the right-most
particle always interact first.

\subsection{A combinatoric formula}

We now show that our rules for constructing diagrams imply that every
expectation value can be written as a sum over all partitions of the initial set
of momenta, where each term in the sum is the product over all subsets forming
the partition, of the sum of all connected diagrams associated to the subset.
That is, let $P = (p_1,\ldots,p_M)$ be a list of momenta, and for any ordered
sublist $s\subset P$ (i.e. any sublist $(p_{j_1},\ldots,p_{j_m})$ with
$j_1<\ldots<j_m$), let us denote by $C(s)$ the sum of all diagrams associated to
$s$ constructed according to the rules above and which are connected. Then we
show that
\beq\label{combinatorics}
	\bra a_{p_1}\cdots a_{p_M}\ket_\beta
	=\sum_{S}C(s_1)\cdots C(s_n)
\eeq
where the sum is over the partitions $S=\{s_1,\ldots,s_n\}$ of $P$:
\beq
	s_1\subset P,\ldots, s_n \subset P,\qquad \bigcup_is_i=P,\qquad s_i\cap
s_j=\emptyset\quad(i\neq j).
\eeq

Let us call ``admissible'' a diagram based on a list $s$ of momenta, which
satisfies our rules for diagram construction. Let us denote by $D_P$ the set of
all admissible diagrams based on $P$, by $D_s^C$ the set of all admissible
connected diagrams based on the ordered sublist $s$ of $P$, and by $\t D_P =
\cup_S D_{s_1}^C\cdots D_{s_n}^C$ the set of all products (juxtapositions) of
connected admissible diagrams based on all partitions $S$ of $P$. In order to
show \eqref{combinatorics}, we only need to show $D_P = \t D_P$. The proof is in
three steps. First, we show that the connected factors in every admissible
diagram are themselves admissible diagrams, based on the list of momenta which
they connect. Second, we show that every admissible diagram based on $P$ is a
product of connected factors based on a partition $S$ of $P$. These first two
points show that $D_P\subset \t D_P$. Last, we show that every product of
connected admissible diagrams based on a partition $S$ of $P$ is an admissible
diagram based on 
$P$. This shows $\t D_P\subset D_P$, proving the equality.

Given a pair of dots that are connected either with \eqref{eq:diagram_rule_conn}
or with \eqref{eq:diagram_rule_disconn}, we will refer to its horizontal
position as the horizontal position of the rightmost member of the pair, and to
its vertical position as the height of the flat horizontal part of the diagram
component \eqref{eq:diagram_rule_conn} or \eqref{eq:diagram_rule_disconn}. Then,
we remark the following.
\begin{rema}\label{remrules}
The set of all diagrams formed using our rules is the set of all diagrams
obtained from the basic components \eqref{eq:diagram_rule_conn} and
\eqref{eq:diagram_rule_disconn}, under the unique additional condition that the
bottom-to-top ordering of the vertical positions of the connections be in
agreement with the right-to-left ordering of their horizontal position.
\end{rema}
Indeed, it is clear that the rules provide diagrams that satisfy the additional
condition of this remark. Further, given a diagram that does satisfy it, we may
scan the rightmost members of the connected pairs from the right to the left; we
observe that under such a scan the connections exactly agree with the rules.

For the first step of the proof of \eqref{combinatorics}, let us consider a connected factor, and the initial dots
that are being connected within the factor. These correspond to an ordered
sublist $s$ of the initial list $P$, and are horizontally aligned and ordered.
Then, according to Remark \ref{remrules}, the bottom-to-top ordering of the
connections within the factor are in agreement with the right-to-left ordering
of the right-members of the connected pairs within the factor; again according
to Remark \ref{remrules}, this implies that a connected factor is an admissible
diagram.

The second step is simply a consequence of the fact that given an admissible
diagram, the relation according to which two elements of $P$ belong to the same
connected factor is an equivalence relation.

The last step is based on the simple observation that within a connected
admissible diagram, it is aways possible to move the connections vertically so
as to space them out, keeping their order the same. Consider a partition $S$ of
$P$ and a product of connected admissible diagrams based on $S$. Let us draw the
product of diagrams with the original dots corresponding to $P$ in ordered
horizontal alignment. Let us scan the rightmost members of the connected pairs
in the full diagram from the right to the left. Going through connected pairs in
a connected factor we observe the correct vertical ordering of the connections.
As we go from a pair in a connected factor to a new pair in a new connected
factor, we impose the correct vertical ordering by moving up all connections of
the new connected factor at or above the level of the new pair. This does not
affect connections in other connected factors. At the end of this process, the
additional condition of Remark \ref{remrules} is satisfied for the full
diagram, hence the diagram is admissible.


\section{Full Counting Statistics} \label{sect:4}

In this section we calculate the full counting statistics \eqref{Fla}, proving
Formula \eqref{cgf}. First, we show that the large $R$ limit of the average in
the formula \eqref{Fla} can be calculated using the continuous Virasoro algebra
introduced in the previous section. With that, it is clear that after taking the
logarithm, only the connected diagrams survive thanks to \eqref{combinatorics}. Then, we show that in the large
$t$ limit, these connected diagrams have a simple dependence on the number of
generators, and resumming the orders of $\lambda$, we end up with the desired
expression \eqref{cgf}.

\subsection{Full counting statistics in terms of Virasoro algebra}

As explained in Section \ref{sect:1}, in the steady state limit, the system
factorises into subsystems described by the Hamiltonians $H^{(j,j+1)}$, so that
expectation values for local fields can be calculated using the steady state
density matrix \eqref{eq:dness}, see Figure \ref{fig:connect}. By rewriting the
product of exponentials in \eqref{Fla} as
\beq\label{porder}
	e^{\ri \lambda Q(t)}e^{-\ri \lambda Q}=e^{\ri \lambda Q+\ri \lambda
\Delta_t Q}e^{-\ri \lambda Q}=\mathcal{P}\exp\left(\ri\int_0^\lambda
d\lambda'\,\Delta_tQ(\lambda')\right),
\eeq
we see that we need to evaluate steady-state averages of $\Delta_tQ=Q(t)-Q$
evolved with $Q$:
\beq\label{evolvQ}
	\Delta_tQ(\lambda'):=e^{\ri \lambda'Q}\Delta_tQe^{-\ri \lambda'Q},
\eeq
and ordered products thereof. The quantity $\Delta_tQ$ was shown in \eqref{qtq}
to depend only on the incoming fields $h^{\text{in}}_j$, for $j=1,\ldots,N$.
Using \eqref{eq:h_chiral}, we can replace these with the chiral fields
$h^{(j,j+1)}$, and write:
\beq \label{eq:DtQ_chiral}
	\Delta_tQ=\sum_{j=1}^N\Delta\alpha_j\int_0^t dx\, h^{(j,j+1)}(-x).
\eeq
Recall that
$\ri[H_0^j,h^{\text{in}}_{j'}(x)]=\delta_{j,j'}\partial_xh^{\text{in}}_j(x)$.
Since also $\ri[H^{(j,j+1)},h^{(j',j'+1)}(-x)] =
\delta_{j,j'}\partial_x\lt(h^{(j',j'+1)}(-x)\rt)$, the action of
$Q=\sum_j\alpha_jH_0^j$ on $\Delta_tQ$ is the same as the action of
$\sum_j\alpha_jH^{(j,j+1)}$. Therefore, in \eqref{evolvQ}, we can make the
following replacement:
\[
	Q\mapsto \sum_{j=1}^N\alpha_jH^{(j,j+1)}.
\]
Hence, the Q-evolution of ``time'' $\lambda$ is the same as a $H$-evolution where
fields are evolved over a time $\lambda\alpha_j$, which depends on the leg the
fields are on. The above results mean that the path-ordered exponential in
\eqref{porder} actually {\em factorizes} amongst the subsystems $H^{(j,j+1)}$,
as does the stationary density matrix \eqref{eq:dness}. Hence the average in
\eqref{Fla} is a sum of terms of similar form. In order to evaluate these terms,
we need to introduce the appropriate algebraic setup.

The expression \eqref{porder} is plagued by UV divergencies coming from local operators $h^{(j,j+1)}(x)$ at coinciding positions. We will use an explicit regularization below.

As is well known (see for instance the book \cite{CFTbook}), the algebraic setup underlying CFT is based on the
Virasoro algebra. This means that we can calculate the expectation value in
\eqref{Fla} using $N$ commuting copies of a Virasoro algebra, $L_n^{(j,j+1)}$,
$n\in\Z$, $j=1,\ldots, N$:
\beq \label{eq:Virasoro_algebra}
[L_m^{(j,j+1)},L_n^{(j,j+1)}]=(m-n)L_{m+n}^{(j,j+1)}+\frac{c}{12}(n^3-n)\delta_{
m+n,0}.
\eeq
There is one copy for the chiral fields in each subsystem $(j,j+1)$, living on
the line segment $[-\tfrac{R}{2},\tfrac{R}{2}]$ with the endpoints identified
(see Figure \ref{fig:connect}). In terms of this algebra, the energy density
operator and the Hamiltonian are given by
\beq\label{eq:HhVir}
  h^{(j,j+1)}(x)=\frac{2\pi}{R^2}\lt(\sum_{n=-\mathcal{N}}^{\mathcal{N}}L_n^{(j,j+1)}e^{2\pi
\ri nx/R} - \frc{c}{24}\rt),\qquad
H^{(j,j+1)}= \frac{2\pi}{R}\lt(L_0^{(j,j+1)}-\frc{c}{24}\rt),
\eeq
and the steady state density matrix takes the following form:
\beq
\rho_{\text{stat}}=\lim_{R\rightarrow\infty}\frak{n}\lt[e^{-\sum_j\beta_j\frac{
2\pi}{R}L_0^{(j, j+1)}}\rt].
\eeq

In \eqref{eq:HhVir} we have used an explicit UV regularization by summing over a finite number of modes $L^{(j,j+1)}_n$. This is equivalent to smearing out the local densities $h^{(j,j+1)}(x)$.

Note that the constant terms in \eqref{eq:HhVir} cancel out and do not play any role in the considerations below. Performing the integral over $x$, we have, from \eqref{eq:DtQ_chiral},
\beq
	\Delta_tQ=\frac{2\pi}{R^2}\sum_{j=1}^N\Delta\alpha_{j}\sum_{n=-\mathcal{N}}^{\mathcal{N}}L_n^{(j,j+1)}\int_{-t}^0dx\,e^{2\pi \ri nx/R}
	=\sum_{j=1}^{N}\Delta\alpha_j\left(\tilde{S}^{j}+S^{j}\right),
\eeq
with
\beq \label{eq:def_S_S-tilde_R}
	\tilde{S}^{j}:=\frac{2\pi t}{R^2}L_0^{(j,j+1)},\qquad
S^{j}:=\frac{2}{R}\sum_{\stackrel{n=-\mathcal{N},}{n\neq 0}}^{\mathcal{N}}L_n^{(j,j+1)}e^{-\ri\pi n
t/R}\frac{\sin\left(\frac{\pi nt}{R}\right)}{n}.
\eeq

The $Q$-evolution in the path-ordered exponential does not affect $\tilde{S}^j$,
since $[Q,\tilde{S}^j]=0$. However, $S^j$ is affected by the action of $Q$.
Using
\[e^{\ri \lambda'\alpha_j\frac{2\pi}{R}L_0^{(j,j+1)}}L_n^{(j,j+1)}e^{-\ri
\lambda'\alpha_j\frac{2\pi}{R}L_0^{(j,j+1)}}=L_n^{(j,j+1)}e^{-\ri
\lambda'\alpha_j\frac{2\pi}{R}n},\]
we can write the $Q$-evolution of a ``time'' $\lambda'$ as a shift of the time
variable appearing in the exponential in \eqref{eq:def_S_S-tilde_R} of twice the
value $\lambda'\alpha_j$, replacing $S^j$ with $S_{\lambda'\alpha_j}^j$ where
\begin{equation}\label{eq:S_tau}
S^j_{\tau}:=\sum_{\stackrel{n=-\mathcal{N},}{n\neq 0}}^{\mathcal{N}}L_{n}^{(j,j+1)}e^{-\ri\pi
n(t+2\tau)/R}\frac{\sin\left(\frac{\pi nt}{R}\right)}{n}.
\end{equation}

Then, the factorized form of the path-ordered exponential in \eqref{porder} is,
after changing variable to $\tau = \lambda'\alpha_j$ for convenience,
\beq\label{eq:pexpprod}
	\mathcal{P}\exp
\lt(\ri\int_0^{\lambda}d\lambda'\,\Delta_tQ(\lambda')\rt)
=\prod_{j=1}^N\mathcal{P}\exp\left(\frc{\ri\Delta\alpha_j}{\alpha_j}\int_{0}^{
\lambda\alpha_j}d\tau\,(\tilde{S}^j+S^j_\tau)\right).
\eeq
Taking the log, \eqref{Fla} gives a sum over the legs, 
\begin{equation}\label{Ff}
F(\lambda)=\sum_{j=1}^Nf(\alpha_j,\Delta\alpha_j,\beta_j,\lambda)
\end{equation}
with
\beq \label{eq:FCS_per_leg}
f(\alpha,\Delta\alpha,\beta,\lambda):=
\lim_{t\rightarrow\infty}\frac{1}{t}\log\lim_{R\rightarrow\infty}\langle\mathcal
{P}\exp\left(\frc{\ri\Delta\alpha}{\alpha}\int_0^{\lambda\alpha}d\tau\,(\tilde{S
}+S_\tau)\right)\rangle_{\beta,R}.
\eeq
Since the form of this expression is identical for each subsystem $(j,j+1)$, we
here and below use a single copy of the Virasoro algebra
\eqref{eq:Virasoro_algebra}, denoted $L_n$, $n\in\Z$, and similarly we drop the
superscripts $j$ on $\t S$ and $S$. Here, the average
$\langle\cdots\rangle_{\beta,R}$ is the trace over a representation of this
single copy of the Virasoro algebra, with density matrix
$\frak{n}\big[\exp(-\beta\tfrac{2\pi}{R} L_0)\big]$:
\[
	\bra\cdots\ket_{\beta,R} := \Tr\lt(\frak{n}\lt[e^{-\beta\tfrac{2\pi}{R}
L_0}\rt]
	\cdots \rt).
\]

Expanding the path-ordered exponential in \eqref{eq:FCS_per_leg} and taking the
average, we get a sum over $m\geq 0$ of
\beq\label{eq:exponential_expanded}
	\lt(\frac{\ri\Delta \alpha}{\alpha}\rt)^m \int_0^{\lambda\alpha}
d\tau_1\cdots\int_0^{\tau_{m-1}} d\tau_m \langle(\tilde{S}+ S_{\tau_{1}})\cdots
(\t S+S_{\tau_m})\rangle_{\beta,R}.
\eeq
In the following we will show that the limit $R\rightarrow\infty$ of these
averages can be calculated using the continuous Virasoro algebra of the previous
section, with the value \eqref{kbeta} of $k$.

%
\subsection{Expectation values in the large $R$ limit}

We now show that the large $R$ limit of a product of Virasoro generators can be
calculated using the continuous algebra of the previous section. We will proceed
by showing first that this is the case for an expectation value of two
generators, and then by induction that it holds for any number of generators.

It turns out that in the large $R$ limit we should take the UV regularization $\mathcal{N}$ to grow proportionally to $R$, thus introducing a fixed energy cutoff $\Lambda$, given by $\mathcal{N}=\Lambda R$.

In order to make the connection to the continuous algebra clear, we will make
use of the following notation,
\begin{equation}
p_i:=\frc{n_i}R,\qquad
a_{p_i}^{(R)}:=\frac{L_{n_i}}{R}\,(1-\delta_{n_i,0}),\qquad
A^{(R)}:=\frac{L_0}{R^2},\qquad\delta^{(R)}(p_i-p_j):=R\delta_{n_i,n_j},
\label{eq:finite_R_algebra}
\end{equation}
where the label $(R)$ indicates that these quantities are defined for finite
system size $R$. Note in particular that according to our definition,
$a_0^{(R)}=0$, which is for convenience.

In \eqref{eq:exponential_expanded}, we see that we need averages of products of
$\t S$ and $S_\tau$. The latter is a sum over $n\neq 0$, the former contains
$L_0$ only. Hence, we are interested in evaluating sums over $n_i\in\Z$ of
averages of products of operators $a_{p_i}^{(R)}$ and operators $A^{(R)}$  (the
condition $n_i\neq0$ is already implemented in the definition of
$a_{p_i}^{(R)}$). We will evaluate these sums, in the large-$R$ limit, as
integrals over continuous momenta $p_i$ of averages of operators $a_{p_i}$ in
the continuous Virasoro algebra developed in the previous section. More
precisely, we will show that the following is true, for every $M\geq0$, every
$\ell\geq 0$ and every continuous function $f$ of the momenta (on the left-hand
side, $p_i=n_i/R$ with $n_i\in\Z$; on the right-hand side, $p_i\in\R$; we use the
notation ${\bf p} = (p_1,\ldots,p_M)$):
\begin{subequations}\label{eq:propsint}
\begin{align}
\lim_{R\rightarrow\infty}\sum_{{\bf p}=-\Lambda}^{\Lambda}f({\bf p})\,\langle a_{p_1}^{(R)}\cdots
a_{p_M}^{(R)}\rangle_{\beta,R} &=\int_{-\Lambda}^{\Lambda} d^Mp\, f({\bf p})\,\langle
a_{p_1}\cdots a_{p_M}\rangle_{\beta}
\label{eq:proponeint}\\
\lim_{R\rightarrow\infty}
\sum_{{\bf p}=-\Lambda}^{\Lambda}f({\bf p})\,
\langle(A^{(R)})^\ell a_{p_1}^{(R)}\cdots a_{p_M}^{(R)}\rangle_{\beta,R}
&=(k(\beta))^\ell
\int_{-\Lambda}^{\Lambda} d^Mp\, f({\bf p})\,
\langle a_{p_1}\cdots a_{p_M}\rangle_{\beta},\label{eq:proptwoint}
\end{align}
\end{subequations}
as well as a similar statements as \eqref{eq:proptwoint}, with the $A^{(R)}$
operators in all possible positions in the average on the left-hand side. On the
right-hand side of \eqref{eq:propsint}, the averages are calculated using the
continuous Virasoro algebra of the previous section with $k=k(\beta)$. Recall
that $k(\beta)$, defined in \eqref{kbeta}, equals $c/(24\beta^2)$. This is the
expectation value of the operator $A^{(R)}$ in the large $R$ limit (see for instance \cite{CFTbook}):
\beq\label{arexp}
	\lim_{R\to\infty} \bra A^{(R)}\ket_{\beta,R} = \frc{c}{24\beta^2}.
\eeq
The statement \eqref{eq:proptwo}, and its relative with different positions of
$A^{(R)}$, can be interpreted by saying that the operator $A^{(R)}$ tends, in
the infinite-$R$ limit, to a central element, which assumes a fixed,
state-dependent  value.

Defining $a_p^{(R)}$ as being zero at $p=0$ is appropriate because it
automatically implements the condition $n\neq 0$ in the sum defining $S_\tau$.
Also, we note that the large-$R$ limit of $L_0/R$ diverges: it is the operator
that measures the energy in the system. The quantity proportional to $L_0$ and
that has a finite large $R$ limit is $A^{(R)}$, which corresponds to the energy
density. 

The induction proof that we present below is an induction on $M$. We will assume
that
\begin{subequations}\label{eq:props}
\begin{align}
\lim_{R\rightarrow\infty} \langle a_{p_1}^{(R)}\cdots
a_{p_M}^{(R)}\rangle_{\beta,R} &\stackrel{\cdot}=\langle a_{p_1}\cdots
a_{p_M}\rangle_{\beta}
\label{eq:propone}\\
\lim_{R\rightarrow\infty}
\langle(A^{(R)})^\ell a_{p_1}^{(R)}\cdots a_{p_M}^{(R)}\rangle_{\beta,R}
&\stackrel{\cdot}=(k(\beta))^\ell
\langle a_{p_1}\cdots a_{p_M}\rangle_{\beta},\label{eq:proptwo}
\end{align}
\end{subequations}
holds for some $M$ and for all $\ell\geq 0$ (as well as similar equations as
\eqref{eq:proptwo} but with the $A^{(R)}$ operators in different positions), and
show that this implies the same for $M+1$. We will also show that this holds for
$M=2$. Here the equations are understood as distributions, in the sense of
\eqref{eq:propsint}, which we represent by the equality symbol``
$\stackrel{\cdot}=$''. In practice, this means that the limits on the left-hand
sides are taken with fixed $p_i$, terms giving zero under integration on the
right-hand sides are neglected, and the distribution relation
\beq\label{limdelta}
	\delta^{(R)}(p_i-p_j) \to \delta(p_i-p_j).
\eeq
is used. The passage from \eqref{eq:props} to \eqref{eq:propsint} is then simply
done by using the relation
\[\sum_{\stackrel{n=-\Lambda R,}{n\neq 0}}^{\Lambda R}\frac{1}{R}\to \int_{-\Lambda}^{\Lambda} dp.\]

For lightness of notation we will omit the explicit integration boundaries below.

\subsubsection{Proof at order 2}

The left-hand side of \eqref{eq:propone} for the case $M=2$ can be written in
terms of Virasoro generators, using the definitions \eqref{eq:finite_R_algebra}
and assuming for now $p_1\neq0$ and $p_2\neq0$:
\[\langle a^{(R)}_{p_1}a^{(R)}_{p_2}\rangle_{\beta,R}=\frac{1}{R^2}\langle
L_{n_1}L_{n_2}\rangle_{\beta,R}=\frac{1}{R^2}\frac{\text{Tr}\left(e^{-\beta\frac
{2\pi}{R}L_0}L_{n_1}L_{n_2}\right)}{\text{Tr}\left(e^{-\beta\frac{2\pi}{R}L_0}
\right)}.\]
Making use of the cyclicity of the trace and the Virasoro algebra, in particular
the exchange relation
\beq \label{eq:Virasoro_trick}
L_ne^{-\frac{2\pi\beta}{R}L_0}=e^{-\frac{2\pi\beta}{R}n}e^{-\frac{2\pi\beta}{R}
L_0}L_n,
\eeq
the result is
\[
	\langle L_{n_1}L_{n_2}\rangle_{\beta,R}=\frac{1}{e^{2\pi\beta
n_2/R}-1}\left((n_2-n_1)\langle
L_{n_1+n_2}\rangle_{\beta,R}+\frac{c}{12}(n_2^3-n_2)\delta_{n_1+n_2,0}\right).\]
Because $\langle L_{n}\rangle_{\beta,R}=0$ for $n\neq0$, we can write the
average of two generators as
\[\langle L_{n_1}L_{n_2}\rangle_{\beta,R}=\frac{1}{e^{2\pi\beta
n_2/R}-1}\left(2n_2\langle
L_0\rangle_{\beta,R}+\frac{c}{12}(n_2^3-n_2)\right)\delta_{n_1+n_2,0},\]
which we can express back in terms of the quantities $A^{(R)}$, $a^{(R)}_p$, $p$
and $\delta^{(R)}(p)$ using \eqref{eq:finite_R_algebra}:
\beq\label{arar}
	\langle
a^{(R)}_{p_1}a^{(R)}_{p_2}\rangle_{\beta,R}=\frac{1}{e^{2\pi\beta
p_2}-1}\left(2p_2\langle
A^{(R)}\rangle_{\beta,R}+\frac{c}{12}\left(p_2^3+\frac{p_2}{R^2}
\right)\right)\delta^{(R)}(p_1+p_2).
\eeq
Taking the limit $R\rightarrow\infty$ and using the delta-function limit
\eqref{limdelta} and the expectation value of $A^{(R)}$ \eqref{arexp}, we find
\[\lim_{R\rightarrow\infty}\langle
a^{(R)}_{p_1}a^{(R)}_{p_2}\rangle_{\beta,R}\stackrel\cdot=\frac{1}{e^{2\pi\beta
p_2}-1}\left(2p_2\frac{c}{24\beta^2}+\frac{c}{12}p_2^3\right)\delta(p_1+p_2).\]
This is in fact true for general $p_1$ and $p_2$, noting that although for
$p_1=p_2=0$ the left-hand side is zero and the right-hand side is not, the
difference does integrate to zero. The right-hand side corresponds to the value
one would obtain by evaluating $\langle a_{p_1}a_{p_2}\rangle_{\beta}$ using the
diagram rules in the previous section for the choice $k=c/(24\beta^2)$. This
shows \eqref{eq:propone} for $M=2$.

Let us now show \eqref{eq:proptwo} for $M=2$. Starting with $\ell=1$, we can
write the insertion of $A^{(R)}$ in terms of the Virasoro generators as
\[\langle
A^{(R)}a^{(R)}_{p_1}a^{(R)}_{p_2}\rangle_{\beta,R}=\frac{1}{R^2}\langle L_0
\,a^{(R)}_{p_1}a^{(R)}_{p_2}\rangle_{\beta,R}.
\]
In general, inserting an extra $L_0$ in an average of an operator $\mathcal{O}$
can be done by differentiating with respect to $\beta$, 
\beq 
\frac{\partial}{\partial\beta}\langle\mathcal{O}\rangle_{\beta,R}=
-\frac{2\pi}{R}\langle L_0\mathcal{O}\rangle_{\beta,R}+\frac{2\pi}{R}\langle
L_0\rangle\langle\mathcal{O}\rangle_{\beta,R}.
\eeq
Hence, the insertion of $A^{(R)}$ is equivalent to the application of a
differential operator:
\beq\label{derar}
	\bra A^{(R)} \Or\ket_{\beta,R}  = \lt(\bra A^{(R)}\ket_{\beta,R} -
\frc1{2\pi R} \frc{\p}{\p \beta}\rt) \bra \Or\ket_{\beta,R}.
\eeq
Specializing to the case $\Or = a^{(R)}_{p_1}a^{(R)}_{p_2}$ and looking at
\eqref{arar}, one can see that the large $R$ limit commutes with the
$\p/\p\beta$ derivative (in \eqref{arar}, a careful analysis is necessary in
order to see this on the average $\bra A^{(R)}\ket_{\beta,R}$, which is beyond
the scope of this paper). Hence the last term on the right-hand side above
vanishes in the limit, which shows \eqref{eq:proptwo} using \eqref{arexp}.
Insertions of $\ell>1$ operators $A^{(R)}$ are obtained similarly by multiple
applications of the differential operator. Furthermore, the fact that the
position of $A^{(R)}$ is unimportant is an immediate consequence of the
commutation relation
\beq\label{aA}
	[a_p^{(R)},A^{(R)}] = \frc{p}R a_p^{(R)} \to 0 \quad\mbox{as}\quad
R\to\infty
\eeq
where the large-$R$ limit is taken inside averages and with fixed $p$, as above.


\subsubsection{The induction step}
In order to prove that \eqref{eq:propone} holds for all values of $M$, we must
now prove the induction step: if \eqref{eq:propone} holds up to some number $M$
of generators, it must also hold for $M+1$ generators.

The finite-$R$ expectation value of $M+1$ operators, again assuming all $p_i$s
different from zero,
\[
	\langle a^{(R)}_{p_1}\cdots
a^{(R)}_{p_{M+1}}\rangle_{\beta,R}=\frac{1}{R^{M+1}}\langle L_{n_1}\cdots
L_{n_{M+1}}\rangle_{\beta,R},
\]
can be calculated in terms of expectation values with $M$ and $M-1$ operators
using cyclicity of the trace and \eqref{eq:Virasoro_trick} in order to bring
$L_{n_{M+1}}$ cyclically through all other operators. This gives
\[\begin{split}
\langle L_{n_1}\cdots L_{n_{M+1}}\rangle&=\frac{1}{e^{2\pi\beta
n_{M+1}/R}-1}\sum_{i=1}^{M}\langle L_{n_1}\cdots[L_{n_{M+1}},L_{n_i}]\cdots
L_{n_M}\rangle\\
&=\frac{1}{e^{2\pi\beta n_{M+1}/R}-1}\sum_{j=1}^{M}\Bigl[(n_{M+1}-n_i)\langle
L_{n_1}\cdots L_{n_i+n_{M+1}}\cdots L_{n_M}\rangle(1-\delta_{n_i+n_{M+1},0})\\
&\quad+\left(2n_{M+1}\langle L_{n_1}\cdots L_0\cdots
L_{n_M}\rangle+\frac{c}{12}(n_{M+1}^3-n_{M+1})\langle L_{n_1}\cdots
\widehat{L_{n_i}}\cdots L_{n_M}\rangle\right)\delta_{n_i+n_{M+1},0}\Bigr],
\end{split}\]
where in the last line we separated the case where the $n_i$ and $n_{M+1}$ add
up to zero (and the hat indicates that the factor is omitted). Using
\eqref{eq:finite_R_algebra}, this can be re-written as
\beq\label{eq:propone_intermediate_result0}
\begin{split}
\langle a^{\tiny{(R)}}_{p_1}\cdots a^{(R)}_{p_{M+1}}\rangle_{\beta,R}
&=\frac{1}{e^{2\pi\beta p_{M+1}}-1}\sum_{i=1}^{M}\Biggl[(p_{M+1}-p_i)\langle
a^{(R)}_{p_1}\cdots a^{(R)}_{p_i+p_{M+1}}\cdots a^{(R)}_{p_M}\rangle_{\beta,R}\\
&\quad+2p_{M+1}\langle a^{(R)}_{p_1}\cdots A^{(R)}\cdots
a^{(R)}_{p_M}\rangle_{\beta,R} \\
&\quad +\frac{c}{12}\lt(p_{M+1}^3 - \frc{p_{M+1}}{R^2}\rt)\langle
a^{(R)}_{p_1}\cdots \widehat{a^{(R)}_{p_{i}}}\cdots
a^{(R)}_{p_M}\rangle_{\beta,R}\,\delta^{(R)}(p_i+p_{M+1})\Biggr]
\end{split}\eeq
and from the induction assumption expressed in \eqref{eq:props}, we can
immediately evaluate the large-$R$ limit:
\begin{multline}
\lim_{R\rightarrow\infty}\langle a^{(R)}_{p_1}\cdots a^{(R)}_{p_{M+1}}\rangle
\stackrel\cdot=\frac{1}{e^{2\pi\beta
p_{M+1}}-1}\sum_{i=1}^{M}\Biggl[(p_{M+1}-p_i)\langle a_{p_1}\cdots
a_{p_i+p_{M+1}}\cdots a_{p_M}\rangle_\beta\\
+\left(2p_{M+1}k(\beta)
+\frac{c}{12}p_{M+1}^3\right)\langle a_{p_1}\cdots \widehat{a_{p_i}}\cdots
a_{p_M}\rangle_\beta\,\delta(p_i+p_{M+1})\Biggr].
\end{multline}
This is exactly the induction step \eqref{eq:iteration} (along with
\eqref{contalg} and \eqref{kbeta}) used in order to evaluate averages in the
continuous Virasoro algebra. Again this holds for general $p_i$s, noting that at
$p_i=0$ the difference between the left- and right-hand sides are terms that
integrate to zero. This shows \eqref{eq:propone} for $M\mapsto M+1$.

In order to show \eqref{eq:proptwo} (and similar equations with the operators
$A^{(R)}$ at different positions), one may use again a derivative argument based
on \eqref{derar}, and the vanishing, in the large-$R$ limit, of the commutator
\eqref{aA}. The fact that the large-$R$ limit commutes with the derivative
operator $\p/\p\beta$ can be seen recursively from
\eqref{eq:propone_intermediate_result0}, using the observation that this fact
was true in the case $M=2$ and using \eqref{derar} (and again, at every step it
is necessary to use this nontrivial fact on the expectation value $\bra
A^{(R)}\ket_{\beta,R}$).


\subsection{Large $t$ limit}

We can now use the results \eqref{eq:propsint} in order to calculate the
expectations that appear in the expansion \eqref{eq:exponential_expanded}.
Writing $S_\tau$ and $\tilde{S}$ in terms of $A^{(R)}$ and $a^{(R)}_p$ following
\eqref{eq:finite_R_algebra}, where $p=n/R$ takes discrete values, we find
\begin{subequations}
	\beq
		\tilde{S}=\frac{2\pi t}{R^2}L_0=2\pi t \,A^{(R)}
	\eeq
	\beq
		S_\tau=\frac{2}{R}\sum_{n\neq 0}L_ne^{-\ri\pi
n(t+2\tau)/R}\frac{\sin\left(\frac{\pi
nt}{R}\right)}{n}=\frc2R\sum_{p}a^{(R)}_pe^{-\ri\pi p(t+2\tau)}\frac{\sin(\pi
pt)}{p}.
	\eeq
\end{subequations}
Equation \eqref{eq:proponeint} then implies that we can express the large $R$
limit of expectation values of products of $S_\tau$ in terms of the continuous
algebra,
\beq\label{eq:largeRaverages}
	\lim_{R\rightarrow\infty}\langle S_{\tau_1}\cdots
S_{\tau_M}\rangle_{\beta,R}=2^M\int d^M{p}\,\langle a_{p_1}\cdots
a_{p_M}\rangle_{\beta}\prod_{i=1}^M e^{-\ri\pi p_i(t+2\tau_i)}\frac{\sin(\pi p_i
t)}{p_i}.
\eeq
Further, Equation \eqref{eq:proptwoint} and its relatives, i.e.~the statement
that in the large-$R$ limit, $A^{(R)}$ becomes central and assumes the fixed
value $k(\beta)$, implies that in the large-$R$ limit, $\t S$ factorizes and can
be treated as the number $2\pi t\,k(\beta)$. Therefore, the large-$R$ limit of
the expectation value in \eqref{eq:FCS_per_leg} can be written as
\begin{multline}\label{pexpap}
\lim_{R\rightarrow\infty}\langle\mathcal{P}\exp\lt(
\frac{\ri\Delta\alpha}{\alpha}\int_0^{\lambda\alpha}d\tau\,(\tilde{S}
+S_\tau)\rt)\rangle_{\beta,R}\\
= e^{2\pi \ri t\, k(\beta)\,\Delta
\alpha\,\lambda}\,\sum_{M=0}^{\infty}\left(\frac{2\ri\Delta\alpha}{\alpha}
\right)^M\int d^M{p}\,\langle a_{p_1}\cdots
a_{p_M}\rangle_{\beta}\prod_{i=1}^{M}\int_{0}^{\tau_{i-1}}d\tau_j\,e^{-\ri\pi
p_i(t+2\tau_i)}\frac{\sin(\pi p_i t)}{p_i},
\end{multline}
with $\tau_0:=\lambda\alpha$.


In order to calculate the large-time cumulant generating function
\eqref{eq:FCS_per_leg}, we take the logarithm of the above expression. Recall
the combinatoric formula \eqref{combinatorics}: an expectation $\langle
a_{p_1}\cdots a_{p_M}\rangle_{\beta}$ is a sum over partitions $S$ of the list
of momenta $(p_1,\ldots,p_M)$, of products over the partition's parts $s\in
S,\;s\subset(p_1,\ldots,p_M)$, of sums $C(s)$ of connected diagrams linking the
momenta in $s$. Then by standard combinatoric arguments, the logarithm of
\eqref{pexpap} is a series formed by the connected averages $\langle
a_{p_1}\cdots a_{p_M}\rangle_{\beta}^{\text{conn}} := C(p_1,\ldots,p_M)$
evaluated by summing over the connected diagrams. From \eqref{eq:FCS_per_leg}
and \eqref{pexpap}, we have
\begin{multline}
f(\alpha,\Delta\alpha,\beta,\lambda)=
2\pi \ri\, k(\beta)\,\Delta \alpha\,\lambda
\\ +\lim_{t\to\infty}\frac{1}{t}
\sum_{M=2}^{\infty}\left(\frac{2\ri\Delta\alpha}{\alpha}\right)^M\int d^M{p}\,\langle a_{p_1}\cdots
a_{p_M}\rangle_{\beta}^{\text{conn}}\prod_{i=1}^{M}\int_0^{\tau_{i-1}}d\tau_i\,
e^{-\ri\pi p_i(t+2\tau_i)}\frac{\sin(\pi p_i t)}{p_i}.\label{fafterlog}
\end{multline}

Every connected diagram $\gamma$ with $M$ momenta contains exactly one overall
delta function $\delta(p_1+\cdots+p_M)$, and can be written in the form
\beq
	G_M^\gamma({\bf p}) \,\delta(p_1+\cdots+p_M),
\eeq
where $G^\gamma_M$, defined on the hyperplane $p_1+\ldots+p_M=0$, is an entire
function of the momenta except for simple poles when seen as a function of the
sum of any subset of the momenta (as is clear from the diagram rules
\eqref{eq:diagram_rule_conn} and \eqref{eq:diagram_rule_disconn}). Hence also
connected averages have this form,
\beq
	\langle a_{p_1}\cdots a_{p_M}\rangle_{\beta}^{\text{conn}}
	= \lt(\sum_\gamma G^\gamma_M({\bf p})\rt) \delta(p_1+\cdots+p_M)
	=: G_M({\bf p}) \,\delta(p_1+\cdots+p_M) \quad (M\geq 2).
\eeq
Taking into account this form, we may evaluate the large-time limit on the
right-hand side of \eqref{fafterlog} using the formula (see e.g. \cite{bernard2012full})
\begin{equation}\label{specialformula}
\lim_{t\to\infty} \frc1 t \int d^M{p}\,g({\bf
p})\,\lt(\prod_{i=1}^M\frac{\sin(\pi p_i  t)}{\pi
p_i}\rt)\,\delta(p_1+\cdots+p_M) =g(0,\ldots,0),
\end{equation}
which gives rise to
\beqa
f(\alpha,\Delta\alpha,\beta,\lambda)&=&
2\pi \ri\, k(\beta)\,\Delta \alpha\,\lambda
 +
\sum_{M=2}^{\infty}\left(\frac{2\pi\ri\Delta\alpha}{\alpha}\right)^M
G_M(0,\ldots,0) \prod_{i=1}^{M}\int_0^{\tau_{i-1}}d\tau_i \n
&=&
2\pi \ri\, k(\beta)\,\Delta \alpha\,\lambda
 +
\sum_{M=2}^{\infty}\frc{\left(2\pi\ri\,\Delta\alpha\,\lambda\right)^M}{M!}
G_M(0,\ldots,0).
\label{fafterlog2}
\eeqa

Hence, we need to evaluate $G_M(0,\ldots,0)$. Although each $G^\gamma_M({\bf
p})$ has poles which make its value at $p_1=\ldots=p_M=0$ ill-defined (the limit
of $G^\gamma_M({\bf p})$ when ${\bf p}\to (0,\ldots,0)$ does not exist), the sum
of all connected diagrams is well defined at zero momenta. The value of
$G_M(0,\ldots,0)$ can be calculated as follows. Consider the operators $\t a_p
:= a_p + k(\beta) \delta(p)$. These satisfy commutation relations with a central
term that does not contain the term linear in $p$, that is $[a_p,a_q] = (p-q) +
\tfrac{c}{12} p^3\,\delta(p+q)$. Hence these operators are explicitly
independent of $\beta$, and they have a nonzero one-point average given by $\bra
\t a_p\ket_\beta = k(\beta)\delta(p)$. Consider the connected averages $\bra \t
a_{p_1}\cdots \t a_{p_M}\ket_\beta^{\text{conn}}$, defined combinatorically as
usual. The insertion of an operator $\t a_0$ can be obtained simply by
differentiating with respect to $\beta$:
\[
	\bra \t a_{p_1}\cdots \t a_{p_M} \t a_0\ket_\beta^{\text{conn}}
	= -\frc1{2\pi} \frc\p{\p \beta}
	\bra \t a_{p_1}\cdots \t a_{p_M} \ket_\beta^{\text{conn}}
\]
for any $M\geq 1$. Since we take connected averages, the shift of the
expectation value is irrelevant whenever the number of operator is greater or
equal to two:  $\bra \t a_{p_1}\cdots \t a_{p_M}\ket_\beta^{\text{conn}} = \bra
a_{p_1}\cdots a_{p_M}\ket_\beta^{\text{conn}}$ for all $M\geq 2$. Hence, we find
the recursion relation
\[
	G_M(0,\ldots,0) = -\frc1{2\pi} \frc{\p}{\p\beta} G_{M-1}(0,\ldots,0)
\]
which holds for all $M\geq 2$ if we define $G_1(0): = k(\beta)$. Using
\eqref{kbeta}, this gives 
\[
	G_M(0,\ldots,0) = \frc{M!\, k(\beta)}{(2\pi \beta)^{M-1}}.
\]
Hence, putting this in \eqref{fafterlog2}, we find
\[
f(\alpha,\Delta\alpha,\beta,\lambda)=2\pi\beta\,k(\beta)
\sum_{M=1}^{\infty}\lt(\frc{\ri\,\Delta\alpha\,\lambda}{\beta }\right)^M
= \frc{2\pi \ri \,\beta\, k(\beta)\,\Delta \alpha\, \lambda}{\beta-\ri\,\Delta
\alpha\,\lambda}.
\]
With \eqref{Ff} and \eqref{kbeta}, this indeed reproduces \eqref{cgf}.

Finally, we observe that the function $f(\alpha,\Delta\alpha,\beta,\lambda)$ is
in fact independent of $\alpha$. Hence, seeing our initial expression
\eqref{porder} as a function of $\alpha_j$ and $\Delta\alpha_j$ in the form
\[
	\Pexp \lt(\ri\int_0^\lambda d\lambda' e^{\ri\lambda' Q(\{\alpha\})}
	(\Delta_t Q)(\{\Delta \alpha\}) e^{-\ri\lambda' Q(\{\alpha\})}\rt),
\]
we conclude that the same result for the full counting statistics is obtained by
setting $\alpha_j=0$ in this expression. But setting $\alpha_j=0$ means setting
$Q=0$, and the exponential becomes simply $e^{i\lambda \Delta_t Q}$, so that we
have shown that the expression \eqref{Fnaive} indeed gives the same result.

\subsection{UV regularization}

In the above derivation, we omitted the explicit momentum cutoff $\Lambda$ in our integrals.
These cutoffs are necessary because the expression \eqref{eq:largeRaverages}, for the large-$R$ limit of expectation
values at finite time $t$, is a multiple momentum integral that is divergent at
large momenta. In fact, as we mentioned, even before taking the large-$R$ limit, at each order in
$\lambda$, the average of $e^{i\lambda Q(t)} e^{-i\lambda Q}$, written in terms
of sums over Virasoro indices $n_i$ of averages of Virasoro generators through
\eqref{eq:pexpprod}, \eqref{eq:def_S_S-tilde_R} and \eqref{eq:S_tau}, is
divergent at large values of $n_i$. These divergencies are expected, as we are
effectively evaluating exponentials of local fields (integrated over finite
regions) in a quantum field theory. Hence, in order to make our calculations
finite, we need a UV regularization. A natural way of regularizing is to modify
the expression of the local fields needed, $h^{(j,j+1)}(x)$: as was done above, one may simply sum
over indices $n\in\Z\cap [-{\cal N},{\cal N}]$ in \eqref{eq:HhVir}, with ${\cal
N}
>0$ finite. Then the finite-$R$ average of $e^{i\lambda Q(t)} e^{-i\lambda Q}$,
which is the average of path-ordered exponential \eqref{eq:pexpprod}, is finite
order by order in $\lambda$. At large $R$, however, we need to make sure that
the UV regularization, which should be an energy, behaves correctly: what needs
to be fixed is not the range of the dimensionless indices $n$, but rather the
range of the momenta $p$. That is, we take ${\cal N}=\Lambda R$ with $\Lambda$ a
fixed quantity with dimension of energy. Then, the regularization translates
into limiting the multiple momentum integral to the range
$p_i\in[-\Lambda,\Lambda]$ in \eqref{eq:largeRaverages}, which is indeed finite.
The large-$t$ limit can then be taken, which kills the integral and constrains the values of the momenta to 0. Hence after
the large-$t$ limit is taken, the result is explicitly independent of $\Lambda$:
the UV regularization can be taken to infinity.

These considerations, and our proof above, gives further confirmation of the claim made in
\cite{bernard2012energy,bernard2013non} that the result for the full counting statistics of the energy flow in
CFT is universal (although it is not a proof and doesn't fully address the problem of irrelevant operators). Indeed, we have here a very different regularization scheme
(perhaps more standard) than that used in \cite{bernard2013non}, yet we find the same result.

The present considerations can also be translated into an intuition as to how the
special steady-state limit, $R\to\infty$ followed by $t\to\infty$, behaves in
terms of the Virasoro modes in CFT. Indeed, as we mentioned, the present
formulation is equivalent to a two-time measurement where the first measurement
is at the connection time. Hence, although the parameter $R$ we used was an
``artefact'' in order to construct the stead-state density matrix, it can be
interpreted as the length of the physical system; and the time $t$ between the
two measurements can be interpreted as the time the system has evolved towards
the steady state. Then, the way we took the $R\to\infty$ limit indicates that
the large-volume physics occurs at very large modes, $n\sim pR$ with fixed
momenta $p$, and the result of the $t\to\infty$ limit indicates that the
large-time limit is dominated by very small values of momenta $p$.


\section{Conclusion} \label{sect:conclu}

We have evaluated the scaled cumulant generating function for the energy
transfer in NECFT on a star graph with temperature imbalances amongst the legs
of the graph, and a simple connection condition at the vertex. We introduced new techniques in order to perform the calculation,
in particular studying a continuous version of the Virasoro algebra and the
associated diagrams. This generalizes the NECFT results of \cite{bernard2012energy,bernard2013non} to the
star-graph configuration, and agrees, in the particular case of unit central
charge, with previously obtained results concerning free bosons (Luttinger
liquids) \cite{mintchev2013luttinger} and free Dirac fermions \cite{mintchev2011non} on star graphs. Our results
further confirm the universality of the NECFT full counting statistics, as well
as the Poisson-process picture underlying long-time energy transfer in NECFT.

There are interesting open problems that the techniques of the present paper may
help address. An immediate question concerns the energy or charge transfer in
the cases of a nontrivial but conformal impurity at the vertex of the graph.
Since impurities are usually described algebraically in CFT, it is possible that
the present Virasoro-algebraic methods may be of use in order to make progress.
Another set of non-equilibrium problems that are of current interest are those
where parameters of a system are suddenly changed and the system then let to
evolve. It is expected that the result at large time is a so-called ``generalized Gibbs
ensemble'', described by a nontrivial density matrix. The techniques
developed here may help in evaluating averages in such density matrices.
Finally, it would also be interesting to generalize the present ideas to
non-conformal (say integrable) situations, for instance with conformal baths and
a non-conformal integrable impurity (like the multi-channel Kondo model).

\appendix

\section{Comparison with other results}\label{sec:comparison}
We compare our results for the energy current with the results obtained in
\cite{mintchev2011non}, where the steady charge and energy currentwere obtained
for a quantum wire junction (both relativistic and nonrelativistic). The quantum
junction is modeled by a star graph where fields living on one of the legs that
are incoming to the vertex have an associated temperature and chemical potential
determined by the leg. The charge and energy exchange are modeled by pointlike
interactions in the vertex, an for comparison with our results, we take these to
be scale-invariant. In the notation of \cite{mintchev2011non} this means the
scattering matrix takes the form
\begin{equation}
\mathbb{S}_{\text{inv}}=\theta(k)\mathbb{U}+\theta(-k)\mathbb{U}^{-1},
\end{equation}
where in \cite{mintchev2011non} the matrix $\mathbb{U}$ is an arbitrary unitary
$N\times N$ matrix, representing the vertex conditions for positive or negative
values of the parameter $k$ (which plays the role of a momentum). We note that
the diagonal elements are reflection at the vertex, and off-diagonal element
$\mathbb{S}_{ij}$ is the transmission amplitude from the $i$-th leg to the
$j$-th leg. This means that the vertex conditions of the present work correspond to
the following choice for $\mathbb{U}$,
\beq
\mathbb{U}_{ij}=\alpha_{i-1}\,\delta_{j,i-1},\qquad
\mathbb{U}_{ij}^{-1}=\alpha_{i}^{-1}\delta_{j,i+1},
\eeq
with the requirement
\beq
\alpha_j^{-1}=\alpha_j^*.
\eeq
The energy flow from the $i$-th leg into the vertex at criticality is given by
\begin{equation}
\mathcal{T}_i(\beta,\mu,\tilde\mu)=\frac{1}{2\pi}\sum_{j=1}^N\left(|\mathbb{U}_{
ij}|^2-\delta_{ij}\right)\frac{1}{\beta_j^2}\left[\text{Li}_2\left(-e^{
-\beta_j\mu_j}\right)+\text{Li}_2\left(-e^{-\beta_j\tilde\mu_j}\right)\right]
\end{equation}
With Li$_2(x)$ known as the dilogarithm (or Spence function Sp$(x)$). Since we
assume that the chemical potentials for the (anti)particles are all equal, we
may take them to zero, and the terms between square brackets become:
\begin{equation}
\left.\text{Li}_2\left(-e^{-\beta_j\mu_j}\right)+\text{Li}_2\left(-e^{
-\beta_j\tilde\mu_j}\right)\right|_{\mu,\tilde\mu=0}=2\text{Li}_2(-1)=-\frac{
2\pi^2}{12}.
\end{equation}
The heat flow for zero chemical potential is given by,
\begin{equation}
\mathcal{T}_i(\beta)=\frac{\pi}{12}\sum_{j=1}^N\left(\delta_{ij}-|\mathbb{U}_{ij
}|^2\right)\frac{1}{\beta_j^2}.
\end{equation}
For our example of sequentially connected CFTs, we have
\begin{equation}
|\mathbb{U}_{ij}|^2=\mathbb{U}^*_{ij}\mathbb{U}_{ij}=\alpha^*_{i-1}\alpha_{i-1}
\delta_{j,i-1}=\alpha^{-1}_{i-1}\alpha_{i-1}\delta_{j,i-1}=\delta_{j,i-1},
\end{equation}
so the heat flow at criticality from the $i$-th leg into the $i+1$-th leg is
given by
\begin{equation}
\mathcal{T}_i(\beta)=\frac{\pi}{12}\left(\frac{1}{\beta_i^2}-\frac{1}{\beta_{i-1
}^2}\right).
\end{equation}
Summing over all legs, and giving each contribution a weight $\Delta\alpha_i$,
it is easy to see that this agrees with our result \eqref{J}, with a choice of
the central charge $c=1$.

\section{An operator interpretation of the change of connection at the vertex of
the graph}\label{sect:appB}

At the time $-t_0$ when the $N$ separate systems are connected as represented in
Figure \ref{fig:connect}, there is a drastic change in the dynamics. Such
changes of boundary or ``impurity'' conditions are often represented in CFT as
insertions of particular local fields at the position of the impurity. Although
this picture is a Euclidean one, and doesn't seem to be directly applicable to
the calculation of quantities in the non-equilibrium steady state, it may be of
use in evaluating overlaps between Hamiltonian eigenstates before and after the
connection.

In the present context, there is a very natural interpretation of the impurity
changing operator. Indeed, before the connection we have $N$ independent chiral
copies of a CFT model, while at the moment of the connection, these copies are
connected to each other in a sequential way. Geometrically, this means that at
the time of the connection there is a branch point. Hence, the impurity
condition changing operator should be the branch-point twist field, a twist field associated to the $\Z_N$ symmetry of any model composed of $N$ independent copies. An elegant derivation of its scaling dimension was found in \cite{CC} in the context of evaluating the entanglement entropy in CFT.

This can be confirmed by analyzing the shift of $L_0$-eigenvalues produced by
the connection, which should be related to the dimension of the branch-point
twist field. This shifts is easily calculated as follows. Consider the total
energy before and after the connection. Before the connection, there are $N$
independent Virasoro algebras, $L_n^{(j)},\,n\in\Z$, representing the single
chiral fields running around each legs of the graph. In each leg $j$, a single
chiral field is used, as is standard in boundary CFT, to represent both
$h_j^{\text{in}}(x)$ and $h_j^{\text{out}}(x)$ and the reflecting boundary
conditions at $x=0$ and $x=R$. After the connection, there is a single Virasoro
algebra $L_n,\,n\in\Z$, representing the single chiral field running around the
whole graph, see the left part of Figure \ref{fig:connect}. This single
chiral field now represents all of the fields $h_j^{\text{in}}(x)$ and
$h_j^{\text{out}}(x)$, for $j=1,\ldots,N$, as well as the special connection at
the vertex of the 
graph. Denoting $\t L_n = L_n  - \frc{c}{24} \delta_{n,0}$ and $\t L_n^{(j)} =
L_n^{(j)}  - \frc{c}{24} \delta_{n,0}$, the energy before the connection is the
sum of those of the $N$ separate systems, each of length $R$:
\[
	\frc{2\pi}{R} \sum_{j=1}^N \t L_{0}^{(j)}.
\]
After the connection, the energy is that of a single system, but of length $NR$,
\[
	\frc{2\pi}{NR} \t L_{0}.
\]
Since the connection only changes the energy by an infinitesimal amount, these
should be equal,
\beq\label{relL0}
	\frc1N L_{0} + \frc{c}{24} \lt(N-\frc1N\rt)
	= \sum_{j=1}^N L_{0}^{(j)}.
\eeq
Hence, the eigenvalues of $\sum_{j=1}^N L_{0}^{(j)}$, in the initial system, are
shifted by $d:=\frc{c}{24} \lt(N-\frc1N\rt)$ as compared to the eigenvalues of
$\frc1N L_{0}$ (the $1/N$ scaling simply represents the change of length). The
number $d$ is indeed the holomorphic dimension of the branch-point twist field
\cite{CC}.

Relation \eqref{relL0} is one of an infinity of relations amongst the Virasoro
generators before and after the connection. These relations can be obtained as
follows. Consider a local right-moving energy density $h_+(z)$, for $z\in
[0,NR]$ running over the chiral paths of all legs of the graph. Before the
connection, $h_+(z)$ is interpreted as representing the chiral fields in all
separate legs, with each leg corresponding to an interval $z\in[jR,(j+1)R]$ of
length $R$, and with periodicity given by identifying the points $z=j$ and
$z=j+R$, for $j=0,\ldots,N-1$. After the connection, the same field $h_+(z)$ is
now interpreted as representing a single chiral field on the interval of length
$NR$, with periodicity where the point 0 is identified with the point $NR$. The
field $h_+(x)$ can then be written in two different ways:
\beqa
	h_+(z) &=& 
	\frc{2\pi}{R^2} \lt(-\frc{c}{24} + \sum_{n\in\Z}
	L_n^{(j)} e^{2\pi \ri nz/R}\rt) \Theta(jR\leq z<(j+1)R) \n
	&=& 
	\frc{2\pi}{R^2} \lt(-\frc{c}{24} + \sum_{n\in\Z}
	L_n e^{2\pi \ri nz/(NR)}\rt) \no
\eeqa
where $\Theta(\cdots)$ is one if the condition $\cdots$ is satisfied, and zero otherwise. The equality between these two expressions gives rise to an infinity of
relations between $L_n$ and $L_n^{(j)}$. In particular, one family of such
relations is
\[
	\frc1N L_{Nk} + \frc{c}{24} \lt(N-\frc1N\rt)\delta_{k,0}
	=\sum_{j=1}^N L_{k}^{(j)},\quad k\in\Z
\]
which generalizes \eqref{relL0}. It is indeed a simple matter to check that both
sides are generators that satisfy the Virasoro algebra commutation relations
with central charge $Nc$.

\bibliographystyle{plain}

\end{document}